\documentclass[%
  aip,
  amsmath,amssymb,
  reprint,           % <- two-column "journal" style
]{revtex4-1}

\usepackage{graphicx}% Include figure files
\usepackage{dcolumn}% Align table columns on decimal point
\usepackage{bm}% bold math

\usepackage[utf8]{inputenc}
\usepackage[T1]{fontenc}
\usepackage{mathptmx}
\usepackage{etoolbox}

\usepackage{xcolor}
\usepackage{algorithm}
\usepackage{algpseudocode}
\usepackage{enumitem}
\usepackage{subcaption}

\usepackage[
    colorlinks=true,
    linkcolor=blue,
    citecolor=blue,
    urlcolor=blue,
    pdfborder={0 0 0},
    bookmarksopen=true,
    bookmarksnumbered=true
]{hyperref}

\makeatletter
\def\@email#1#2{%
 \endgroup
 \patchcmd{\titleblock@produce}
  {\frontmatter@RRAPformat}
  {\frontmatter@RRAPformat{\produce@RRAP{*#1\href{mailto:#2}{#2}}}\frontmatter@RRAPformat}
  {}{}
}%
\makeatother

\newif\ifblinded
\blindedfalse

\ifblinded
  
\else
  
\fi

\makeatletter
\renewcommand{\refname}{REFERENCES} % change "References" title
\makeatother

\makeatletter
\renewcommand{\@makecaption}[2]{%
  \small                     % keep AIP caption size
  \setlength{\parindent}{0pt}%
  \vskip\abovecaptionskip
  \noindent#1. #2\par        % normal paragraph => justified text
  \vskip\belowcaptionskip
}
\makeatother

\begin{document}

\preprint{AIP/123-QED} % optional; change or remove as needed

\title{Flow-Aware Ellipsoidal Filtration for Persistent Homology of Recurrent Signals}

% --- Authors and affiliations (REVTeX style) ---
\author{O~B.~Eryilmaz}
\email{obe851@student.bham.ac.uk}
\affiliation{%
School of Computer Science, University of Birmingham, Birmingham B15 2TT, United Kingdom%
}%

\author{C.~Katar}
\affiliation{%
Electrics and Electronics Department, Turkish-German University, Istanbul 34820, Turkey%
}%

\author{M.~A.~Little}
\affiliation{%
School of Computer Science, University of Birmingham, Birmingham B15 2TT, United Kingdom%
}%

\date{February 25, 2026}

\begin{abstract} Recurrent signals give rise to trajectories that repeatedly return close to earlier states in state space. Many analysis methods therefore require a principled notion of similarity between states. In practice, a recurrence threshold sets the scale of the neighbourhood used to define when two states are considered close. Close returns can also support topology-preserving denoising in state space, aiming to reduce noise while preserving the trajectory's structure, which classical denoising methods may distort. The effectiveness of both denoising and recurrence analysis therefore depends critically on how these neighbourhoods are modelled and scaled. 

This work introduces a flow-aware ellipsoidal filtration for persistent homology based on a spatio--temporal covariance construction that estimates local flow geometry from both temporal and spatial neighbours. Unlike isotropic constructions based on balls (e.g.\ the Vietoris--Rips filtration), the proposed method assigns an ellipsoid to each point, with orientation and axis lengths determined by local flow variances. When a dominant $H_1$ feature reflects the recurrent loop structure, its persistence interval provides a data-driven scale selection. Across the considered experiments, flow-aware ellipsoidal neighbourhoods improve topology-preserving denoising and first-recurrence-time estimation relative to the Vietoris--Rips filtration. Overall, the results indicate that persistent homology can be more informative for dynamical systems when domain knowledge is used to incorporate anisotropy. 
\end{abstract} 
\maketitle % ========================= % Main text % ========================= 
\textbf{
Persistent homology is increasingly used to analyse time-series data from dynamical systems. However, most applications rely on isotropic distance-based neighbourhoods, which may not capture the local flow structure of a trajectory. In this paper, we construct flow-aware ellipsoidal neighbourhoods for each data point using local spatio–temporal covariance. Persistent homology is then used to guide the scaling of these neighbourhoods, which are applied to topology-preserving denoising and close-return detection. The resulting neighbourhoods more faithfully preserve trajectory geometry and improve recurrence-time estimation in the presence of narrow bottlenecks.}

\section{Introduction} 

Recurrent behaviour is common in nonlinear systems and forms the basis of recurrence-based analysis methods \citep{webberRecurrenceQuantificationAnalysis2015, marwanTrendsRecurrenceAnalysis2023}. To use these returns in practice, one must specify when two states should be considered close, which is typically done by choosing a recurrence threshold that sets the neighbourhood scale for proximity \citep{schinkelSelectionRecurrenceThreshold2008,zouComplexNetworkApproaches2019,liOptimizingSelforganizedTopology2025}. This choice is important because the same neighbourhood definition can be used both to quantify recurrence structure \citep{webberRecurrenceQuantificationAnalysis2015} and to denoise in state space by local averaging \citep{robinsonTopologicalLowPassFilter2016}, while aiming to preserve the trajectory’s overall shape. By contrast, standard denoising methods may distort this structure when applied without reference to state-space geometry. For these reasons, both recurrence analysis and topology-preserving denoising depend critically on how neighbourhoods are defined and how their scale is selected from data. 

Persistent homology \citep{edelsbrunnerTopologicalPersistenceSimplification2002, Edelsbrunner_Persistent_Homology_a_Survey,carlssonTopologicalDataAnalysis2021}, a tool from applied and computational topology, provides a multiscale description of the geometric and topological structure of these sampled trajectories. Instead of analysing the data at a single resolution, it records how topological features emerge and disappear as a neighbourhood scale parameter increases. This process is governed by a \emph{filtration}, which specifies how local neighbourhoods around data points grow and intersect across scales. Common constructions include the alpha, \v{C}ech, and Vietoris--Rips filtrations. In practice, the Vietoris--Rips filtration is frequently used because of its computational simplicity, defining connectivity through intersections of isotropic balls under a chosen distance metric. 

Time-series analysis with persistent homology largely relies on sliding-window embeddings to convert one-dimensional signals into high-dimensional point clouds. We focus on time-series analysis via point-cloud representations; readers seeking further examples of TDA applications may consult the DONUT repository \citep{DONUT}. Perea and Harer \cite{pereaSlidingWindowsPersistence2015} established a theoretical link between the most persistent \(H_1\) class, signal periodicity, and the choice of window length and embedding dimension. Gakhar and Perea \cite{gakharSlidingWindowPersistence2024} further showed that sliding-window embeddings of quasi-periodic signals are dense in an \(n\)-torus, which can be characterised via persistent homology. Vejdemo-Johansson \cite{vejdemo-johanssonCohomologicalLearningPeriodic2015} used persistent cohomology to construct circular coordinates from sliding-window embeddings, enabling periodicity detection, dimensionality reduction, and denoising; this was later generalised to toroidal coordinates for quasi-periodic signals \citep{scoccolaToroidalCoordinatesDecorrelating2023}.

More broadly, persistent homology has been applied to dynamical-systems data to study delay-coordinate reconstructions of chaotic attractors \citep{maleticPersistentTopologicalFeatures2016}, to guide embedding-delay selection (SToPS) \citep{tanSelectingEmbeddingDelays2023}, and to assess whether learned models reproduce reference dynamics from sampled trajectories \citep{tanGradingYourModels2021}. Complementing these point-cloud-based approaches, cubical homology on phase-portrait images yields topological descriptors (e.g.\ Betti curves and persistence-based features) for automated analysis and classification of dynamical behaviours \citep{hussainshahTopologicalDataAnalysis2025}. Most of the above constructions use Vietoris--Rips filtrations under an ambient Euclidean metric; under the manifold assumption, this motivates alternatives that modify the neighbourhood relation, such as density-aware intrinsic metrics (Fermat distance) \citep{fernandezIntrinsicPersistentHomology2023} or temporally informed anisotropic neighbourhoods.

A further motivation for anisotropic neighbourhoods comes from the manifold hypothesis, which posits that high-dimensional data may lie near a lower-dimensional manifold and motivates algorithmic tests of such structure \citep{feffermanTestingManifoldHypothesis2016}. In this setting, isotropic balls can be a crude local model, since the data may vary primarily along tangent directions \citep{marchettiIntrinsicDimensionalityFermi2025}. Breiding \textit{et al.}\ proposed replacing $\varepsilon$-balls by ellipsoids aligned with estimated tangent spaces (obtained, when polynomial equations are available, from Jacobian information), and reported longer persistence bars in their examples \citep{breidingLearningAlgebraicVarieties2018}. Complementing this, Kali\v{s}nik and Le\v{s}nik proved that unions of tangent-elongated ellipsoids can deformation retract to a manifold under suitable sampling assumptions, with improved sample-density requirements over ball thickenings \citep{kalisnikFindingHomologyManifolds2024}.  Canova \textit{et al.}\ \cite{canovaPersistentHomologyEllipsoids2025} define an ellipsoid complex for point clouds sampled from manifolds. Each point is assigned an ellipsoid whose orientation is given by local \emph{principal component analysis} (PCA) on \(k\)-nearest neighbours and whose axis ratios are fixed globally; simplices are added when the corresponding ellipsoids intersect. These ellipsoidal constructions were developed for manifold reconstruction or shape classification, where persistence diagrams are converted into feature vectors for classification tasks. Neighbourhoods are defined using purely spatial criteria, without incorporating temporal ordering of the samples.

While these methods operate on static point clouds, our work addresses point clouds sampled from dynamical systems. It is motivated by the search for adaptive, data-driven neighbourhoods for topological denoising of quasi-periodic signals \citep{robinsonTopologicalLowPassFilter2016}, rather than relying on fixed \(k\)-nearest-neighbour (\(k\)-NN) constructions. In contrast to these spatial constructions, in our earlier work \citep{eryilmazEllipsoidalFiltrationTopological2025}, we introduced an \emph{adaptive ellipsoidal filtration} that incorporates local flow information by shaping anisotropic neighbourhoods according to signal gradients, thereby enabling flow-aware persistent homology for recurrent dynamics. However, in that formulation the local neighbourhood was defined purely by temporal windowing, which limited its ability to capture variability in close returns. 

In the present study, this framework is extended in several ways. First, the adaptive ellipsoid construction is refined by defining spatio–temporal neighbourhoods that jointly consider local spatial proximity and short-term temporal evolution, allowing more accurate estimation of local flow geometry. Second, the method is applied to real accelerometer recordings of recurrent motion, validating its practical relevance. Third, ellipsoidal filtrations are compared with Vietoris–Rips and Fermat filtrations to evaluate differences in topological representation. Finally, the utility of the resulting neighbourhoods for recurrence-time estimation is assessed, and improved performance over Vietoris–Rips-based approaches is demonstrated. The main contributions are: 
\begin{itemize} 
\item A flow-aware geometric interpretation of point-cloud density for recurrent dynamics is introduced. For uniformly sampled trajectories, sparse regions of the point cloud correspond to fast-flow segments, and dense regions to slow-flow segments. Around each point, a flow-aware ellipsoidal neighbourhood region is attached, whose size and orientation reflect the local flow, and points are connected when their ellipsoids intersect. In the experiments considered here, the ellipsoidal filtration recovers the main loop in attractors with bottlenecks (i.e.\ narrow pinch regions of the loop), whereas Vietoris–Rips and Fermat filtrations tend to split this into two loops (Section~\ref{subsec:filtration_behaviour_exp}).

\item It is shown that the ellipsoidal filtration provides a principled and interpretable guideline for neighbourhood selection in topological low-pass filtering of recurrent signals \citep{robinsonTopologicalLowPassFilter2016}. We assume a suitable embedding in which recurrent behaviour manifests as loop structures in state space, so that significant $H_1$ features correspond to this recurrence. The death scale of the most persistent $H_1$ class is used as a conservative upper bound for neighbourhood selection, ensuring that the dominant loop structure remains observable. Empirically, we find that the variance-adaptive nature of the flow-aware ellipsoidal neighbourhoods can preserve the underlying geometric structure even slightly beyond this bound. Compared with $k$-NN and Vietoris--Rips-based isotropic neighbourhoods, the proposed flow-aware ellipsoids more consistently preserve fine-scale structures across bottlenecks (Sections~\ref{sec:DenoisingRecurrentSignal}--\ref{sec:accDenoising}).

\item Recurrence period density methods \citep{littleExploitingNonlinearRecurrence2007} are revisited, and a topologically grounded way to choose recurrence threshold is provided. Vietoris–Rips-based isotropic balls are compared with flow-aware ellipsoids, with both scaled using guidance from persistent homology, specifically the scale of the dominant $H_1$ class. The resulting flow-aware ellipsoidal recurrence neighbourhoods improve first-return time estimation compared with Vietoris–Rips-based constructions (Section~\ref{sec:recurrence-time}). 
\end{itemize} 

The paper is structured as follows. Section~\ref{sec:background} reviews background on recurrent signals and persistent homology. Section~\ref{sec:problem_statement} formulates the problem for temporally sampled dynamical data with observational noise and motivates flow-aware neighbourhoods. Section~\ref{sec:methods} presents the construction of the flow-aware ellipsoidal filtration. Section~\ref{sec:experiments_and_results} reports experiments on dense–sparse phase-space structure, denoising of synthetic and 3D accelerometer signals, and recurrence-time estimation. Section~\ref{sec:discussion} discusses implications, limitations and future directions, and Section~\ref{sec:conclusion_and_future_work} concludes.

\section{Background}
\label{sec:background}
This section introduces recurrence in state space and summarises persistent homology, establishing the notation used throughout the paper.
\subsection{Recurrent Signals}
\label{sec:background-recurrence}

Recurrent signals are characterised by repeated patterns, which appear as returns to previously visited regions in state space. These patterns often reflect the presence of underlying deterministic dynamics.

We model the evolution of such signals as a continuous flow governed by a dynamical system. Let \( \mathbf{u}(t) \in \mathbb{R}^d \) denote the latent state at time \( t \), evolving according to
\begin{equation}
\frac{d\mathbf{u}}{dt} = f(\mathbf{u}(t), t),
\label{eq:dudt}
\end{equation}
where \( f : \mathbb{R}^d \times \mathbb{R} \to \mathbb{R}^d \) is a smooth, nonlinear function. Given an initial condition, the solution traces a trajectory in state space that represents the system’s temporal evolution.

In practice, we observe only a finite sequence of noisy, discrete-time samples:
\begin{equation}
\mathbf{v}_i = \mathbf{u}(t_i) + \mathbf{n}(t_i), \quad \text{with } t_i = t_0 + i \Delta t,
\label{eq:obsnoise}
\end{equation}
where \( \mathbf{n}(t_i) \) denotes additive noise. The resulting point cloud approximates the underlying trajectory and enables geometric and topological analysis.

We say that the system exhibits a \emph{recurrence} at time \( t \) if there exists a delay \( \delta_t > 0 \) such that
\begin{equation}
\label{eq:first_return_condition}
\begin{aligned}
\mathbf{u}(t+\delta_t) &\in \mathbf{B}\big(\mathbf{u}(t), r\big), \\
\mathbf{u}(t+\tau) &\notin \mathbf{B}\big(\mathbf{u}(t), r\big)
\quad \text{for all } 0 < \tau < \delta_t .
\end{aligned}
\end{equation}

where \( \mathbf{B}(\mathbf{u}(t), r) \) is the closed ball of radius \( r > 0 \) centred at \( \mathbf{u}(t) \). The value \( \delta_t \) is interpreted as the approximate recurrence time, indicating the system’s first return to a neighbourhood of its earlier state \citep{kantzNonlinearTimeSeries2003}.

Recurrence time \( \delta_t \) depends strongly on the neighbourhood radius \( r \). If \( r \) is too large, nearly all states may appear recurrent; if too small, most true recurrences may be missed. Choosing a suitable threshold is therefore essential, and various strategies have been proposed in the literature \citep{schinkelSelectionRecurrenceThreshold2008,liOptimizingSelforganizedTopology2025}.

This formulation naturally includes both periodic and aperiodic dynamics despite these two seeming mutually exclusive. In the case of exact periodicity, recurrence times \( \delta_t \) remain constant, and recurrence occurs at fixed intervals even as \( r \to 0 \). In contrast, real-world signals are generally aperiodic, and \( \delta_t \) varies over time. In such cases, recurrence is approximate and highly dependent on the choice of neighbourhood size. The study of these variations—known as recurrence time statistics—has a long history in nonlinear time-series analysis \citep{kantzNonlinearTimeSeries2003}.

Understanding and analysing these time-varying recurrence structures requires methods that are robust to noise and sensitive to local geometry. Persistent homology provides a principled, multi-scale framework for this purpose, motivating our use of topological methods in the study of recurrent signals. In this study, we propose a data-driven approach that defines ellipsoidal neighbourhoods to determine the recurrence threshold based on local variance. The scale of these neighbourhoods is determined using persistent homology to capture the global topological structure of the signal.

\subsection{Persistent Homology}
\label{sec:persistent_homology}
Persistent homology is a foundational technique in topological data analysis (TDA) \citep{carlssonTopologicalDataAnalysis2021,chazalIntroductionTopologicalData2021}. It identifies and tracks topological features—connected components, loops, and higher-dimensional voids—across a scale parameter \(\varepsilon\).

The workflow has two steps. First, construct a \emph{filtration}, i.e. a nested family of complexes \(\{K_\varepsilon\}_{\varepsilon\ge 0}\) built from the data by a proximity rule (e.g. Vietoris–Rips: add a simplex when all pairwise distances of its vertices are \(\le \varepsilon\)). As \(\varepsilon\) increases, the complex becomes more connected and features appear and disappear. Second, compute homology on each \(K_\varepsilon\) and record for every feature its \emph{birth} scale \(b\) and \emph{death} scale \(d\). The \emph{persistence} (or lifetime) is
\(
\ell = d - b,
\)
which quantifies the feature’s prominence across scales.

\begin{figure*}[t]
  \centering
  \includegraphics[width=0.95\linewidth,
                   trim=0 70 0 10,clip]{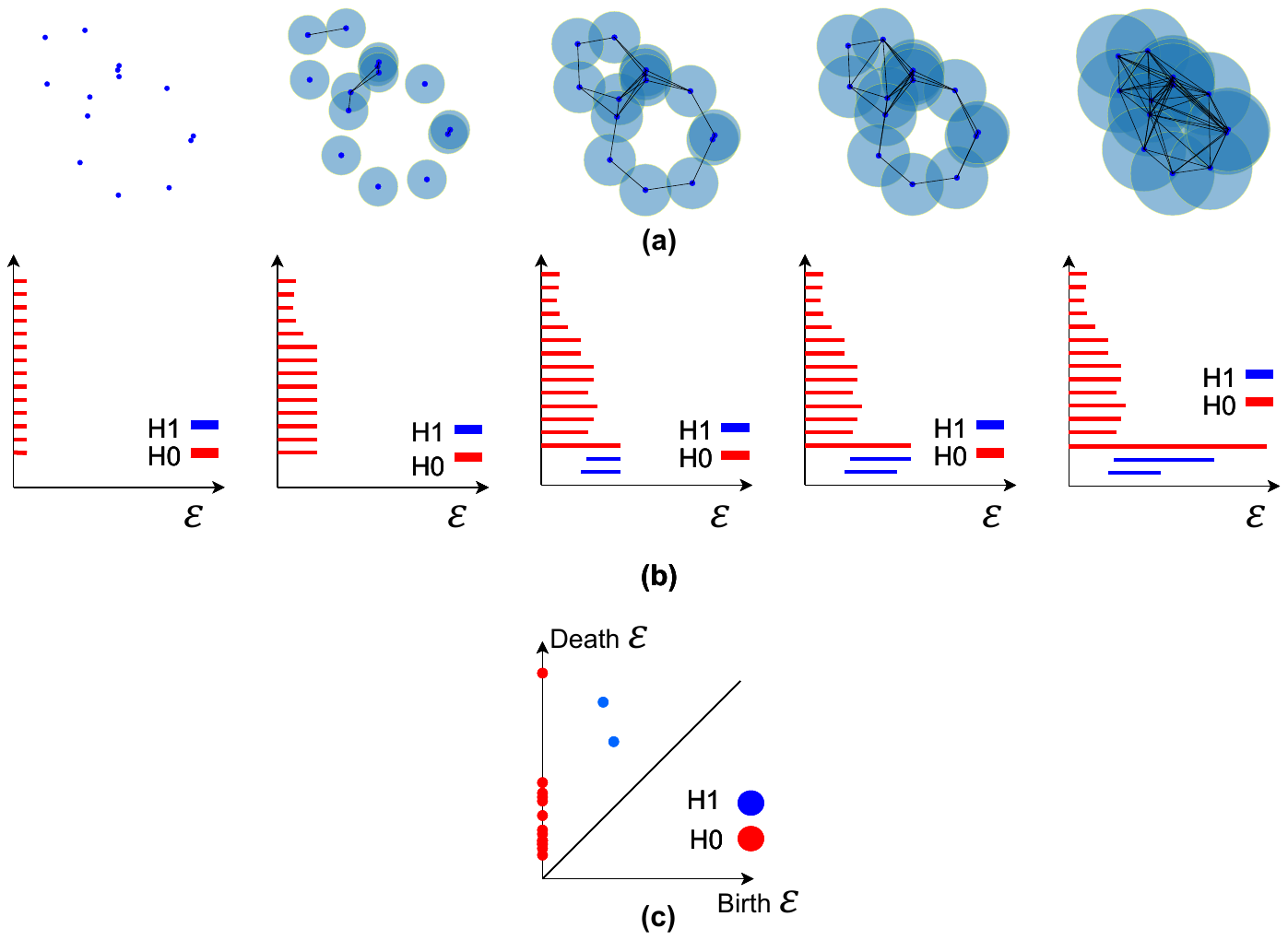}
  \caption{Illustration with a Vietoris–Rips (VR) filtration: 
  (a) data and connections at increasing \(\varepsilon\); 
  (b) barcode showing \(H_0\) (components) and \(H_1\) (loops) as horizontal intervals; 
  (c) persistence diagram (PD), mapping each feature to its birth–death point \((b,c)\).
  The dominant \(H_1\) class is farthest from the diagonal.}
  \label{fig:vr_pd}
\end{figure*}

Formally, for a finite sample \(\mathbb{X}\subset\mathbb{R}^n\), a filtration
\begin{equation}
K_{\varepsilon_1} \subseteq K_{\varepsilon_2} \subseteq \cdots \subseteq K_{\varepsilon_m}
\end{equation}
is obtained by increasing \(\varepsilon\). Exact recovery of the topology of the underlying space is often impossible with finite, noisy data, but persistent homology yields a robust multi-scale summary; this procedure is summarised in Fig.~\ref{fig:vr_pd}.

\section{Problem Statement}
\label{sec:problem_statement}

Point clouds are often modelled as samples from a distribution concentrated near a geometric object \(X \subset \mathbb{R}^n\), with noise controlling the spread around \(X\)~\citep{carlssonTopologyData2009}.
Under this view, common filtration techniques such as Vietoris–Rips and Čech are used to infer the Betti numbers of \(X\) from a finite sample \(\mathbb{X}\) (see Section~\ref{sec:persistent_homology}).
Because these constructions relate points solely through pairwise spatial distances, they impose a uniform, isotropic threshold to define neighbourhoods and build simplicial complexes.
However, in point clouds derived from time-series data, the relationships between samples reflect both spatial and temporal proximity.

In this work, a different setting is considered in which the observed point cloud \( \mathbb{V} \subset \mathbb{R}^n \) is sampled at uniform time intervals from a dynamical process with observation noise.
It is assumed that there exists an unobserved latent trajectory \( \mathbf{u}(t) \) evolving according to the smooth flow introduced in Section~\ref{sec:background-recurrence}, governed by Eq.~(\ref{eq:dudt}), and that noisy samples are observed according to Eq.~(\ref{eq:obsnoise}).

Thus, the point cloud \( \mathbb{V} = \{ \mathbf{v}(t_i) \}_{i=1}^N \) represents a time-parametrised dynamical process rather than a static geometric object, with its spatial density reflecting the local speed of the flow—denser in slow regions and sparser in fast ones.
This anisotropic induced spatial density invalidates the assumption that uniform spatial distances between points define meaningful neighbourhoods, and standard isotropic filtrations may therefore fail to recover the true topological structure of the trajectory \( \mathbf{u}(t) \).

The core problem addressed in this paper is to construct neighbourhoods and filtrations that respect the local dynamics of the flow. 
Specifically, the aim is to approximate the manifold of the underlying trajectory using locally estimated, PCA-aligned ellipsoids instead of isotropic balls; to select a data-driven scale, based on the dominant \(H_1\) class, that captures global structure while preserving local detail; and to exploit these flow-aware neighbourhoods both for topology-preserving denoising and for improved recurrence-time estimation.
\section{Methods}
\label{sec:methods}
This section describes how the proposed ellipsoidal filtration is constructed to incorporate local flow information from recurrent signals. The goal is to replace isotropic neighbourhoods, which ignore the direction and speed of evolution, with anisotropic ellipsoids that adapt to the local geometry of the trajectory. The construction proceeds in four main steps: (i) estimating local covariance from spatio–temporal neighbourhoods, (ii) defining adaptive ellipsoids aligned with local flow directions, (iii) detecting intersections between ellipsoids to determine connectivity, and (iv) constructing a simplicial complex suitable for persistent homology.

\subsection{Local Covariance Estimation}
\label{sec:local_covariance}

The first step is to estimate how the signal varies locally around each sample. 
Given a point cloud representation of the signal 
\( X = \{\mathbf{x}_1, \mathbf{x}_2, \dots, \mathbf{x}_n\} \subset \mathbb{R}^d \),
each point \( \mathbf{x}_i \) represents the system's state at a discrete time step. 
To capture both the temporal evolution and the recurrence structure, two types of local neighbourhoods are used:
(i) an \emph{index-based (temporal) neighbourhood}, which captures short-term evolution in time, and
(ii) a \emph{state-space (spatial) neighbourhood}, which identifies recurrent states that are close in state-space but possibly far apart in time.

The temporal neighbourhood is defined by a symmetric window around each point:
\begin{equation}
\mathcal{T}_i = \{ \mathbf{x}_j \mid |j - i| \leq \tau \},
\label{eq:temporal_neighbourhood}
\end{equation}
where \( \tau \in \mathbb{N} \) is the half-width of the window. This neighbourhood describes the short segment of the trajectory immediately surrounding the point \( \mathbf{x}_i \).

The spatial neighbourhood consists of the \(k\)-nearest neighbours in Euclidean distance:
\begin{equation}    
\mathcal{S}_i = \text{$k$-nearest neighbours of } \mathbf{x}_i \text{ in state space}.
\end{equation}
This captures recurrent states that return to the same region of the attractor, even when they occur far apart in time.

The combined neighbourhood is given by
\begin{equation} 
\mathcal{N}_i = \mathcal{T}_i \cup \mathcal{S}_i,
\end{equation}
which unites local temporal continuity with spatial recurrence.

Spatial neighbours capture the variance arising from close returns, while temporal neighbours encode the local direction and rate of change. As the trajectory typically revisits the same regions many times, we choose \(|\mathcal{S}_i| > |\mathcal{T}_i|\). Accordingly, the parameters \(k\) and \(\tau\) are selected to promote stable local tangent estimates, taking into account the sampling density and noise level of the data.

The local geometry around each point is characterised by the empirical covariance matrix
\begin{equation}    
\Sigma_i = \frac{1}{|\mathcal{N}_i|} 
\sum_{\mathbf{x}_j \in \mathcal{N}_i} 
(\mathbf{x}_j - \mathbf{x}_i)(\mathbf{x}_j - \mathbf{x}_i)^\top.
\end{equation}
This covariance matrix encodes both the direction and magnitude of local variation, effectively capturing how the trajectory is oriented and stretched in state-space. The eigenvectors of \( \Sigma_i \) indicate the dominant flow directions, while the eigenvalues reflect how rapidly the signal varies along each direction.

\subsection{Adaptive Ellipsoids from Local Covariance}
\label{sec:adaptive_ellipsoids}

Once the local covariance is obtained, it is used to construct an anisotropic neighbourhood in the form of an ellipsoid. Let
\begin{equation}
    \Sigma_i = Q_i \Lambda_i Q_i^\top
\end{equation}
be the eigendecomposition of the covariance matrix, where \( Q_i \) contains the orthonormal eigenvectors and \( \Lambda_i = \mathrm{diag}(\lambda_{i,1}, \dots, \lambda_{i,d}) \) the eigenvalues, ordered such that \( \lambda_{i,1} \geq \cdots \geq \lambda_{i,d} > 0 \).

The adaptive ellipsoid centred at \( \mathbf{x}_i \) and scale \( \varepsilon > 0 \) is defined by
\begin{equation}    
E_i(\varepsilon) 
= \left\{ \mathbf{x} \in \mathbb{R}^d 
\;\middle|\; 
(\mathbf{x} - \mathbf{x}_i)^\top \Sigma_i^{-1} (\mathbf{x} - \mathbf{x}_i) \leq \varepsilon^2
\right\}.
\end{equation}

Each ellipsoid is aligned with the local flow directions (eigenvectors) and stretched according to the local variance (eigenvalues). The semi-axis lengths are proportional to \( \varepsilon \sqrt{\lambda_{i,j}} \), meaning that as \( \varepsilon \) increases, each ellipsoid expands uniformly while preserving its anisotropy. This allows the filtration to respect the local geometry of the trajectory at every scale.

\subsection{Intersection of Adaptive Ellipsoids}
\label{sec:ellipsoid_intersection}

To construct a topological complex, we must determine whether two ellipsoids intersect. If two ellipsoids overlap at scale \( \varepsilon \), the corresponding points are connected by an edge. 

Let \(A = \Sigma_i / \varepsilon^2\), \(B = \Sigma_j / \varepsilon^2\), and \(v = \mathbf{x}_i - \mathbf{x}_j\). Following the formulation in \citet{Gilitschenski_Ellipsoid_Intersection}, define
\begin{equation}
K(S) = 1 - v^\top 
\left(
\frac{1}{1-S} B^{-1} + \frac{1}{S} A^{-1}
\right)^{-1} 
v, 
\quad S \in (0,1).
\label{eq:K_function}
\end{equation}
The two ellipsoids \(E_i(\varepsilon)\) and \(E_j(\varepsilon)\) intersect if and only if
\begin{equation}
\min_{S \in (0,1)} K(S) \geq 0.
\end{equation}

Since \(K(S)\) is convex on \((0,1)\), this minimisation is a one-dimensional convex optimisation problem. In practice, we employ a golden-section search with an early termination criterion: if a negative value of \(K(S)\) is encountered during the search, the two ellipsoids are declared non-intersecting and the algorithm immediately proceeds to the next pair.

\subsection{Ellipsoid Complex}
\label{sec:ellipsoid_complex}

Given the family of ellipsoids \(\{E_i(\varepsilon)\}_{i=1}^n\) centred at the points of the dataset, we define the \emph{ellipsoid complex} at scale \(\varepsilon\) as
\begin{equation}
\mathcal{K}_\varepsilon = 
\left\{ 
\sigma \subseteq X 
\;\middle|\;
E_i(\varepsilon) \cap E_j(\varepsilon) \neq \emptyset 
\ \text{for all distinct}\ i,j \in \sigma
\right\}.
\end{equation}
Intuitively, two points are connected if their local uncertainty regions—represented by ellipsoids—overlap. Higher-order simplices are added whenever all pairwise intersections among their vertices are non-empty. This definition yields a \emph{flag complex} whose 1-skeleton (graph) is determined by ellipsoid intersections.

As the scale parameter \(\varepsilon\) increases, the ellipsoids grow and the complex becomes more connected, generating a nested sequence
\(
\mathcal{K}_{\varepsilon_1} \subseteq \mathcal{K}_{\varepsilon_2} \subseteq \cdots,
\)
which defines a valid filtration for persistent homology analysis. In this way, ellipsoidal filtration generalises the Vietoris–Rips construction by replacing isotropic balls with anisotropic, flow-aware ellipsoids.

\subsection{Algorithm for Building the Ellipsoid Complex}
\label{sec:algorithm}

Algorithm~\ref{alg:ellipsoid_complex} summarises the computational procedure for building the ellipsoid complex at a given scale. For each pair of points, the algorithm checks whether their adaptive ellipsoids intersect according to the convex criterion in Eq.~\eqref{eq:K_function}. The resulting edge list defines the 1-skeleton, from which the flag complex is constructed. The flag complex construction and persistent homology were implemented with the \texttt{SimplexTree} data structure from the \textsc{GUDHI} library \citep{mariaGudhiLibrarySimplicial2014}.

\begin{algorithm}[H]
\caption{Ellipsoid Complex Construction at Scale $\varepsilon$}
\label{alg:ellipsoid_complex}
\begin{algorithmic}[1]
\Require Point cloud $X = \{\mathbf{x}_1, \dots, \mathbf{x}_n\}$, covariance matrices $\{\Sigma_i\}$, scale $\varepsilon$
\State Initialise edge set $E \gets \emptyset$
\For{$i = 1$ to $n$}
    \For{$j = i+1$ to $n$}
        \State $A \gets \Sigma_i / \varepsilon^2$, \quad $B \gets \Sigma_j / \varepsilon^2$, \quad $v \gets \mathbf{x}_j - \mathbf{x}_i$
        \State Compute $\min_{S \in (0,1)} K(S)$ using golden-section search with early termination
        \If{$\min K(S) \geq 0$}
            \State Add edge $(i,j)$ to $E$
        \EndIf
    \EndFor
\EndFor
\State Form the flag complex $\mathcal{K}_\varepsilon$ from the 1-skeleton $(X,E)$
\State \Return $\mathcal{K}_\varepsilon$
\end{algorithmic}
\end{algorithm}

\begin{figure*}[t]
  \centering
  \includegraphics[width=0.95\linewidth,
                   trim=150 100 150 100,clip]{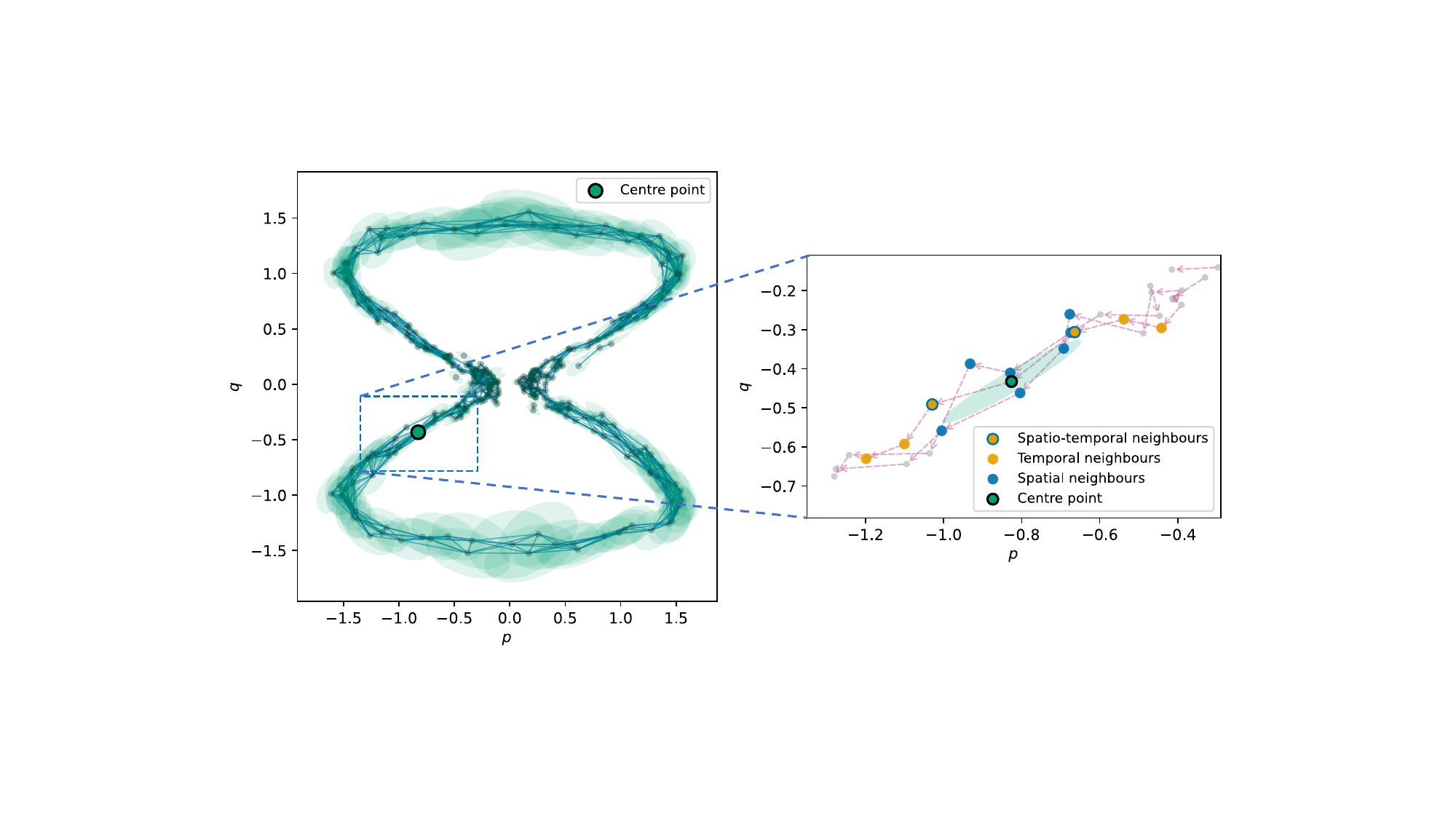}
  \caption{Flow-aware ellipsoids constructed from combined spatial and temporal
  neighbourhoods. Left: recurrent trajectory in the $(p,q)$-plane with
  ellipsoids placed along the orbit at a fixed scale. Right: zoomed view of the centre point, together with its temporal
  neighbours, spatial neighbours, and spatio–temporal neighbours (points that
  are close both in space and in time), as indicated in the legend. The
  ellipsoid is fitted by PCA to the union of the temporal and spatial neighbour
  sets, so that all ellipsoids share the same scale parameter but adapt their
  size and orientation to the local flow geometry.}
  \label{fig:creating_ellipsoids}
\end{figure*}

Repeating this process over an increasing sequence of scales
\(0 < \varepsilon_1 < \varepsilon_2 < \cdots\) produces the filtration
\(\{\mathcal{K}_{\varepsilon_t}\}\) used for persistent homology computation.
This filtration captures how topological features of the data—such as loops and
connected components—emerge and disappear as neighbourhoods expand along the
flow’s geometry. Fig.~\ref{fig:creating_ellipsoids} illustrates this
construction: the left panel shows ellipsoids placed along a recurrent
trajectory at a fixed scale, while the right panel zooms into a single centre
point and displays its temporal, spatial, and spatio–temporal neighbours. The
ellipsoid at the centre point is obtained by applying PCA to the union of the
temporal and spatial neighbour sets.

\noindent\textbf{Code availability.}
The implementation and analysis scripts will be released in a public repository after the paper is accepted.

\section{Experiments and Results}
\label{sec:experiments_and_results}
\subsection{Filtration behaviour with dense and sparse phase–space regions}
\label{subsec:filtration_behaviour_exp}
\begin{figure*}[t]
    \centering

    \begin{subfigure}{0.9\textwidth}
        \centering
        \includegraphics[width=\linewidth]{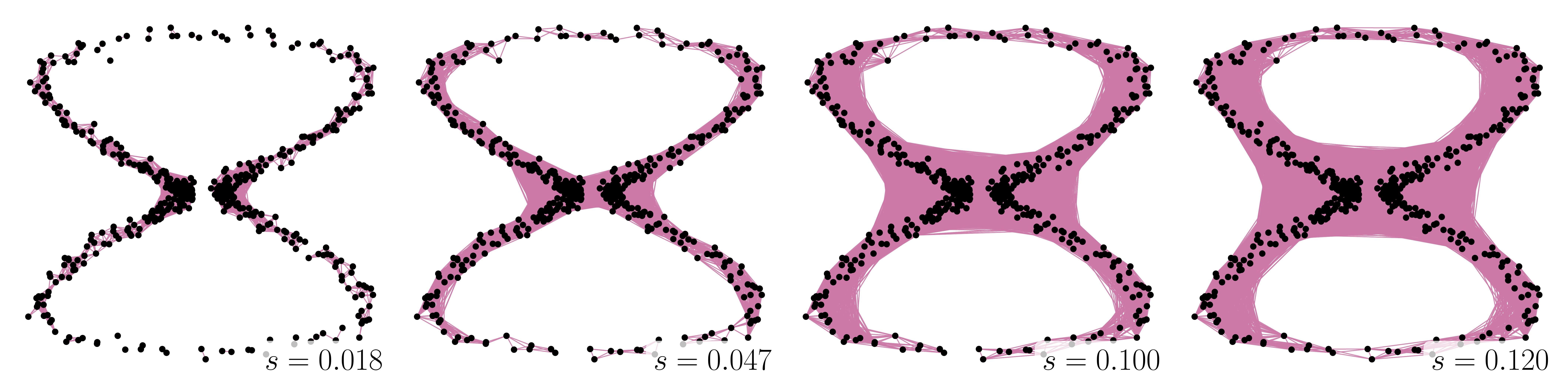}
        \subcaption{}
        \label{fig:fermat}
    \end{subfigure}

    \medskip

    \begin{subfigure}{0.9\textwidth}
        \centering
        \includegraphics[width=\linewidth]{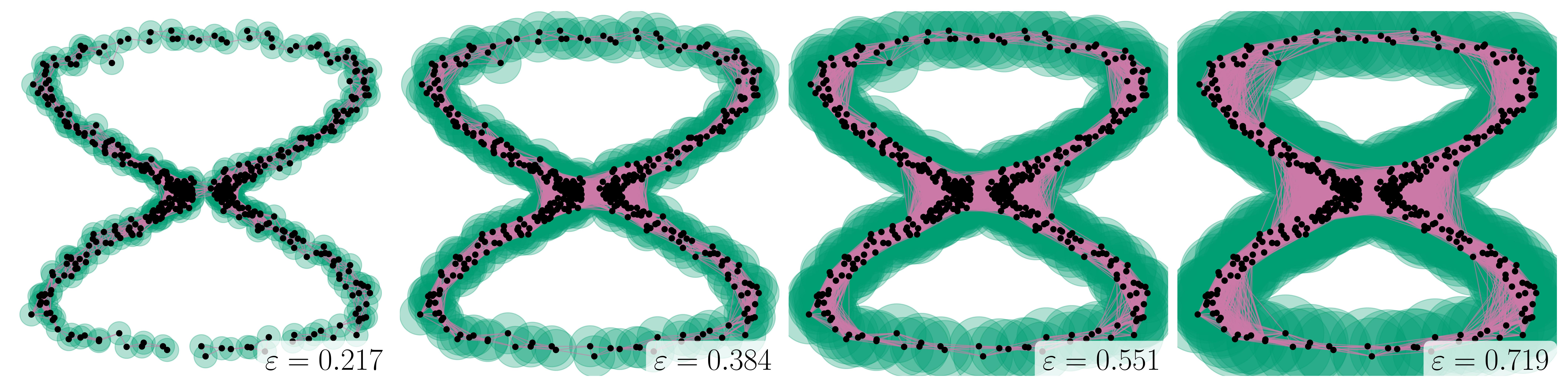}
        \subcaption{}
        \label{fig:vr}
    \end{subfigure}

    \medskip

    \begin{subfigure}{0.9\textwidth}
        \centering
        \includegraphics[width=\linewidth]{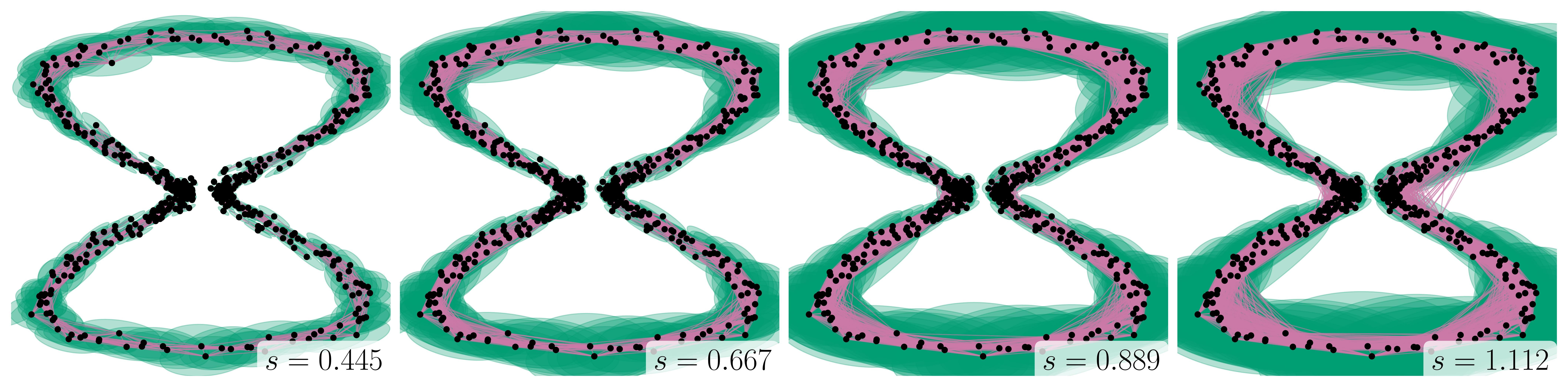}
        \subcaption{}
        \label{fig:ellipsoid}
    \end{subfigure}

    \caption{Comparison of three filtrations on the same dataset generated by the Hamiltonian system in~\eqref{eq:hamiltonian}. Each row shows a $1\times4$ grid across increasing scale values. Pink line segments indicate edges added when the corresponding filtration condition is satisfied; green regions visualise the associated neighbourhoods. In all panels, $p$ is the horizontal axis and $q$ is the vertical axis.
(a) Fermat-distance-based filtration: edges are added when $d_{\text{Fermat}}(x_i,x_j)\le \varepsilon$.
(b) Vietoris--Rips (VR) filtration: edges are added when $\lVert x_i-x_j\rVert \le \varepsilon$.
(c) Ellipsoidal filtration: edges are added when the ellipsoidal neighbourhoods intersect, $E_i(s)\cap E_j(s)\neq\varnothing$.
Here, $\varepsilon$ is the distance threshold for the Fermat-distance-based and VR filtrations, while $s$ scales the ellipsoidal neighbourhoods.}
\label{fig:filtration_comparison}

\end{figure*}

We generate a two-dimensional trajectory from a Hamiltonian system with a non-linear potential,
\begin{equation}
\label{eq:hamiltonian}
\dot{q} = p, 
\qquad 
\dot{p} = -\big(q^r - q - \epsilon \omega \sin(\omega q)\big),
\end{equation}
with parameters $r=5$, $\epsilon=0.39$, and $\omega=3$. The system is discretised using the Störmer--Verlet method with time step $h=0.08$ over a total integration time $T=37.22$, and initial conditions $(q_0,p_0)=(0.5,1.0)$. Gaussian noise is added independently to both coordinates to mimic measurement uncertainty. The resulting trajectory, shown in state space $(p,q)$ in Fig.~\ref{fig:filtration_comparison}, is used to test the filtrations. The same dynamical system and numerical discretisation are used for the method demonstration in Fig.~\ref{fig:creating_ellipsoids}, with a slightly shorter integration time $T=27.22$.

Although the trajectory is sampled at uniform time intervals, the induced point cloud in state space is highly non-uniform: slow dynamics yield dense regions, whereas fast dynamics leave sparse ones. When temporal information is not incorporated, a purely spatial filtration may connect nearby points that belong to different phases of the trajectory. This tendency to connect geometrically close but dynamically distinct states makes the system a useful test case for evaluating how different filtrations recover the trajectory’s topology from its point cloud representation. Importantly, the same fast–slow structure also arises in practice, for example in 3D accelerometer recordings of cyclic human motion.

We compare three filtrations: ellipsoidal, Vietoris--Rips (VR), and Fermat distance. 
VR depends only on Euclidean distance, whereas the Fermat distance incorporates sampling density through an intrinsic path-based metric \citep{fernandezIntrinsicPersistentHomology2023}. 
For a finite set of points $\mathbb{X}_n$, the sample Fermat distance between $x,y \in \mathbb{X}_n$ is defined as
\begin{equation}    
d_{\mathbb{X}_n,p}(x,y) = \inf_{\gamma} \sum_{i=0}^{r} \lvert x_{i+1} - x_i \rvert^p,
\end{equation}
where the infimum ranges over all discrete paths 
$\gamma = (x_0, x_1, \dots, x_{r+1})$ with endpoints $x_0 = x$ and $x_{r+1} = y$, 
and intermediate points $x_1, \dots, x_r \in \mathbb{X}_n$. We set $p=2$ in this experiment.

Fig.~\ref{fig:filtration_comparison} compares the three filtrations across increasing neighbourhood scales. The aim is to recover the global loop that reflects the system’s recurrence, despite the narrow bottleneck in the middle. With Fermat distance (Fig.~\ref{fig:filtration_comparison}\subref{fig:fermat}), dense regions connect quickly, but the bottleneck is bridged before sparse regions are included, preventing loop recovery. Vietoris–Rips (Fig.~\ref{fig:filtration_comparison}\subref{fig:vr}) also connects across the bottleneck too early and likewise fails to recover the loop. In contrast, the ellipsoidal filtration (Fig.~\ref{fig:filtration_comparison}\subref{fig:ellipsoid}) adapts neighbourhoods to local flow direction and speed. This allows the loop to appear at smaller scales while the bottleneck remains separated in the leftmost plot, giving a more accurate representation of the underlying topology.

In this dataset, neighbourhoods based only on density or Euclidean distance did not recover the target loop, because the scale parameter does not reflect the local flow structure. Ellipsoidal filtration combines temporal information with spatial proximity, enabling the correct loop structure to be recovered even in the presence of bottlenecks. Although this dataset is synthetic, similar structures are common in real-world signals, such as accelerometer data from running and cycling (see Section~\ref{sec:accDenoising}). The following experiments build on this result, showing how appropriate neighbourhood selection is crucial for denoising recurrent signals and estimating recurrence times.

\subsection{Denoising a recurrent signal}
\label{sec:DenoisingRecurrentSignal}
We evaluate whether flow-aware neighbourhoods improve the denoising of recurrent signals. Building on earlier evaluations of topological filters \citep{eryilmazEllipsoidalFiltrationTopological2025}, we replicate the experimental setting and make two changes: (i) a refined ellipsoidal filtration based on spatio–temporal PCA, and (ii) an updated synthetic dataset with the chirp phase defined as the integral of the instantaneous frequency, \(\phi(t)=2\pi\!\int_0^t f(s)\,ds\).

The synthetic signal is designed to expose two representative challenges: a linearly increasing frequency and a narrow bottleneck in the trajectory. We define
\begin{subequations}
\label{eq:chirp_xy}
\begin{align}
x(t) &= 10\cos\!\big(\phi(t)\big), \\
y(t) &= 2\sin\!\big(\phi(t)\big)\,S\!\big(x(t)\big),
\end{align}
\end{subequations}
for \(t\in[0,2]\) with \(n = 500\) uniform samples. The instantaneous frequency is
\begin{equation}
\label{eq:inst_freq}
f(t)
  = f_{\mathrm{start}}
    + \frac{f_{\mathrm{end}} - f_{\mathrm{start}}}{t_{\max}}\,t.
\end{equation}
Here \((f_{\mathrm{start}}, f_{\mathrm{end}}) = (1,10)\,\mathrm{Hz}\). The corresponding chirp phase is
\begin{equation}
\phi(t)=2\pi\!\int_0^t f(s)\,ds
=2\pi\!\left(f_{\mathrm{start}}t+\frac{(f_{\mathrm{end}}-f_{\mathrm{start}})}{2t_{\max}}t^{2}\right).
\end{equation}
Anisotropy in \(y\) is introduced by
\begin{equation}
S(x)=1-0.9\exp\!\left(-\tfrac12\left(\frac{x}{\max_{t}|x(t)|}\right)^2\right).
\end{equation}

We add independent, zero-mean Gaussian noise to each coordinate, with axis-wise variances \(\sigma_x^2=P_x/\mathrm{SNR}\) and \(\sigma_y^2=P_y/\mathrm{SNR}\), where \(P_x\) and \(P_y\) are the mean-square powers of the clean \(x\) and \(y\) signals. This yields a bivariate Gaussian noise with diagonal covariance, horizontally elongated \((\sigma_x>\sigma_y)\) and temporally white; the overall 2D SNR equals the target value. Under severe noise, the most persistent \(H_1\) scale can be unreliable; here we therefore treat the \(H_1\) death time as a guide rather than a guarantee.

We compare five strategies: (i) a Vietoris--Rips-based (isotropic, “spherical”) topological filter, (ii) an ellipsoidal (flow-aware) topological filter built using the spatio--temporal PCA construction as described in Section~\ref{sec:methods}, (iii) a \(k\)-NN topological filter with \(k=20\), (iv) a fixed 20-sample moving average, and (v) an adaptive moving average whose window length is set from an FFT-based frequency estimate (Nyquist rule). For Vietoris--Rips and ellipsoidal topological filters, the neighbourhood scale is chosen by the death time of the most persistent \(H_1\) class.

\begin{figure*}[t]
  \centering

  % Left block: x(t)
  \begin{subfigure}[t]{0.48\textwidth}
    \centering
    \includegraphics[width=\linewidth,
                     height=0.42\textheight,
                     keepaspectratio,
                     page=1]{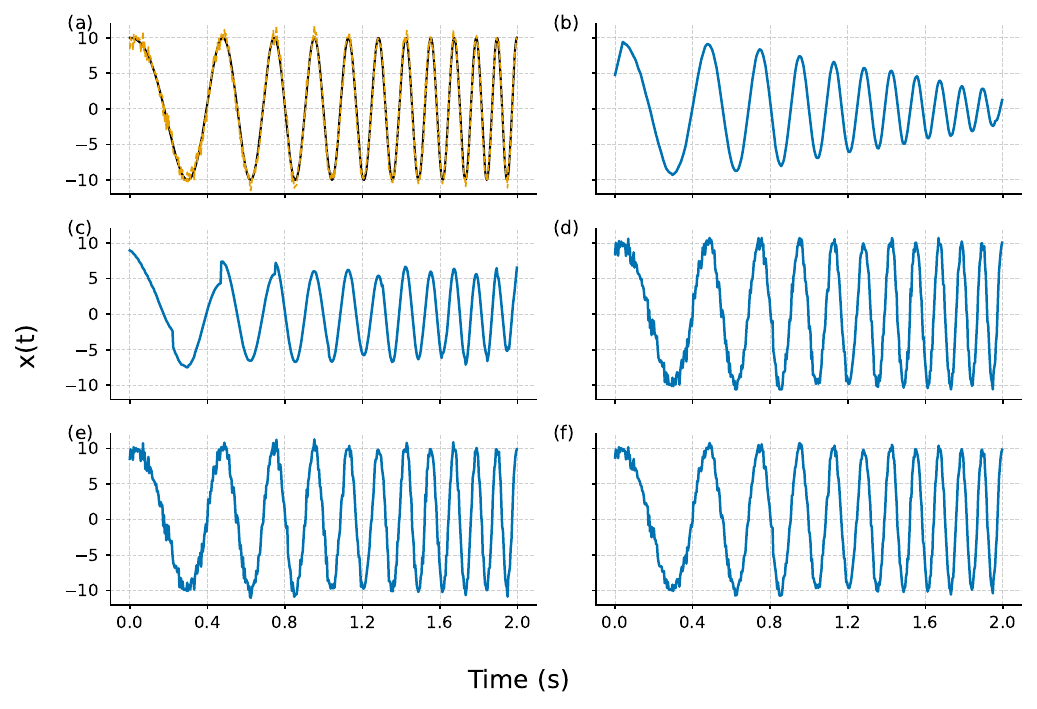}
    
    \label{fig:filtered_signals:x}
  \end{subfigure}
  \hfill
  % Right block: y(t)
  \begin{subfigure}[t]{0.48\textwidth}
    \centering
    \includegraphics[width=\linewidth,
                     height=0.42\textheight,
                     keepaspectratio,
                     page=2]{figures/filtered_signals2026-02-24.pdf}
    
    \label{fig:filtered_signals:y}
  \end{subfigure}

  \caption{Comparison of denoising filters for the two-component signal at SNR = 20~dB.
  Left: high-amplitude component \(x(t)\); right: low-amplitude component \(y(t)\).
  In both panels, (a) original (clean) and noisy signals, (b) moving average,
  (c) adaptive moving average, (d) \(k\)-NN topological filter, (e) spherical topological filter,
  and (f) ellipsoidal topological filter.}
  \label{fig:filtered_signals}
\end{figure*}

Even when the noise level is modest (20\,dB), the methods differ markedly in how well they preserve the input structure (Fig.~\ref{fig:filtered_signals}). The noisy input exhibits a clear frequency rise and a narrow low-amplitude bottleneck. The \emph{fixed} 20-sample moving average is the most clearly destructive: once the period drops below the window length, the filter averages across opposite phases of the oscillation and the waveform is driven towards zero in the middle segment; fine peak details disappear. The \emph{adaptive} moving average is not sufficiently responsive to preserve the signal structure. As the frequency increases, it fails to adjust its window size accordingly, leading to reduced peak-to-peak amplitude, rounded peaks, and loss of fine details.

Topological filters behave differently for the low-amplitude component shown in the right panel of Fig.~\ref{fig:filtered_signals}. The \(k\)-NN filter preserves oscillations over large portions of the record, yet in the bottleneck its fixed-size neighbourhood spans opposing directions of the trajectory, causing local attenuation and shape distortion. The VR (spherical) filter is more severely affected because isotropic balls cut across the thin geometry; cross-bottleneck averaging reduces peak-to-peak amplitude in the low-amplitude region. By contrast, the ellipsoidal filter aligns neighbourhoods with the local flow direction and speed, avoiding cross-bottleneck mixing; amplitude and phase are preserved across the record, including the bottleneck.

\begin{figure}[t]
    \centering
    \includegraphics[width=0.8\linewidth]{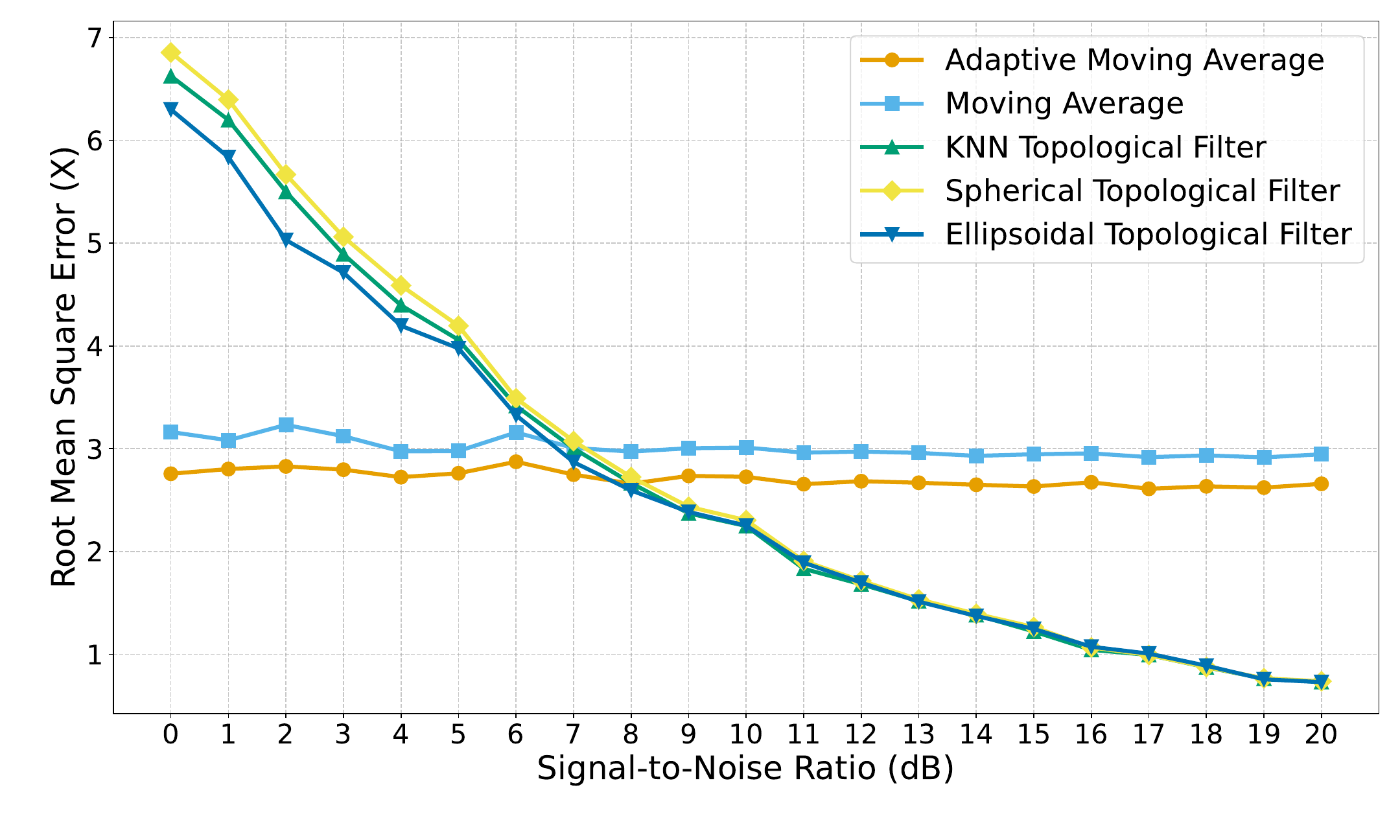}

    \vspace{0.2em}

    \includegraphics[width=0.8\linewidth]{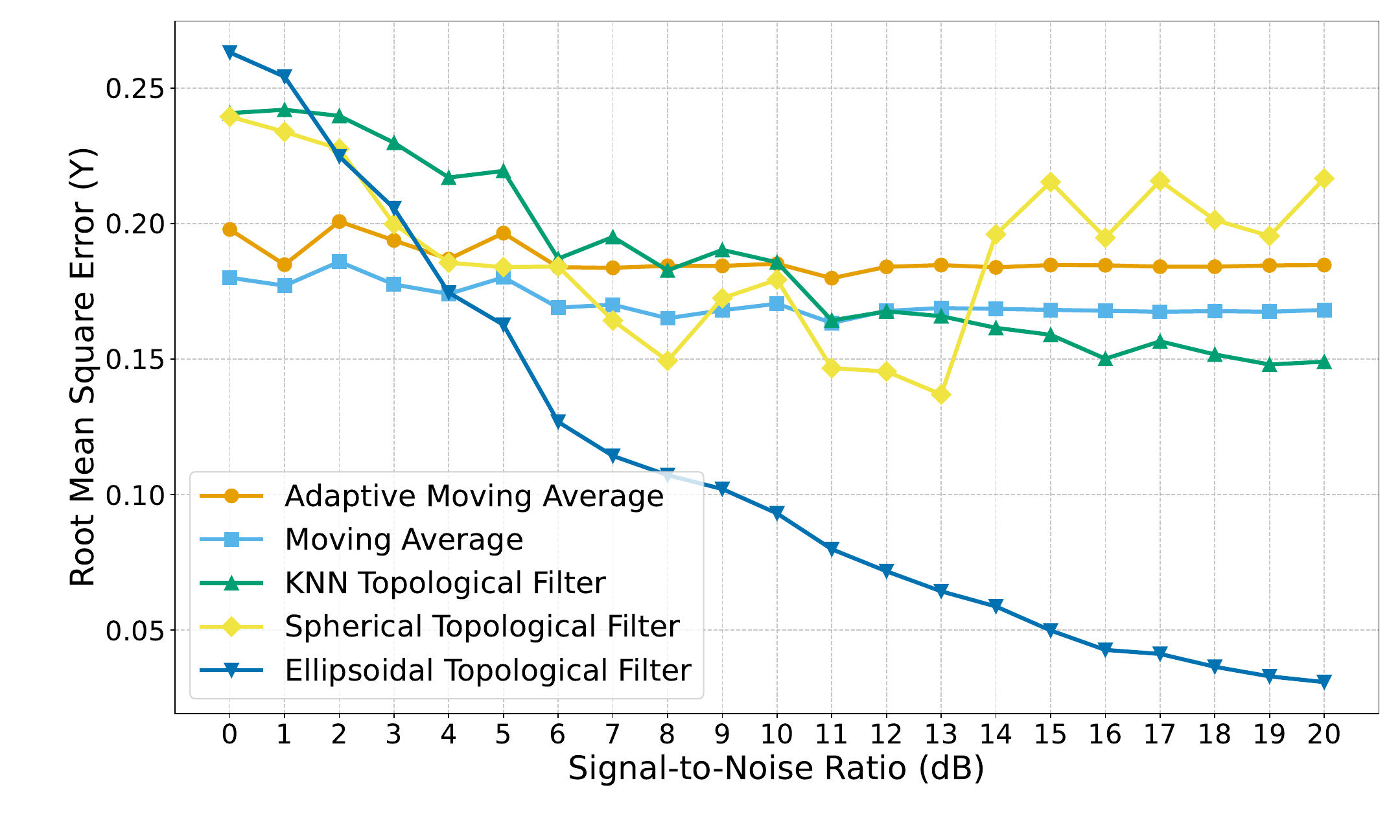}

    \caption{RMSE versus SNR for the chirp signal components:
top panel shows the high-amplitude component $x(t)$, and bottom panel the low-amplitude component $y(t)$.
}
    \label{fig:snr_rmse}
\end{figure}

Fig.~\ref{fig:snr_rmse} reports root-mean-square error (RMSE) versus SNR for \(x(t)\) and \(y(t)\). For \(x(t)\), the moving-average baselines can yield lower RMSE at low SNRs because aggressive smoothing suppresses noise; however, their error quickly plateaus as SNR increases. The three topological filters improve steadily with SNR and overtake the baselines from moderate SNRs upwards, with broadly similar performance among the topological methods on \(x(t)\). For \(y(t)\), the contrast is stronger: the ellipsoidal filter achieves the lowest RMSE over the mid- to high-SNR range, while \(k\)-NN and spherical VR also surpass the time-domain baselines once the SNR is moderate. This behaviour reflects a trade-off: time-domain smoothing reduces noise but also attenuates peak-to-peak structure, so its RMSE gains saturate, whereas topology-based neighbourhoods preserve structure and therefore continue to improve as noise decreases.

\subsection{3D accelerometer data denoising}
\label{sec:accDenoising}

\begin{figure}[t]
  \centering

  \begin{subfigure}{0.4\linewidth}
    \centering
    \includegraphics[width=\linewidth]{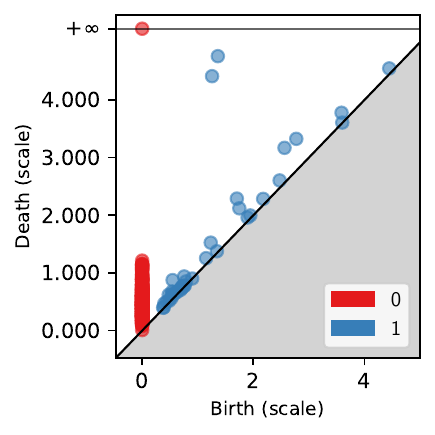}
    
  \end{subfigure}
  \hspace{0.3em}
  \begin{subfigure}{0.4\linewidth}
    \centering
    \includegraphics[width=\linewidth]{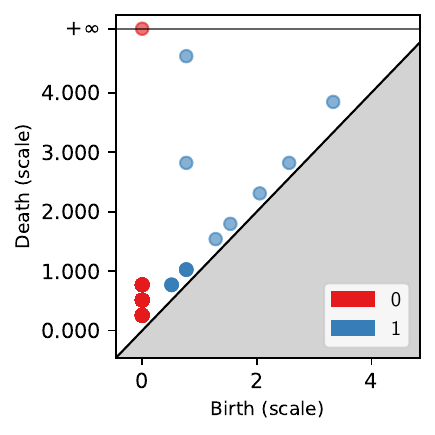}
  \end{subfigure}

  \caption{
Persistence diagrams computed on the 3D accelerometer signal using Vietoris--Rips (spherical, left) and flow-aware ellipsoidal (right) filtrations. 
Under the Vietoris--Rips filtration, two $H_1$ loops with similar persistence appear close to one another, whereas the ellipsoidal filtration yields a clearer separation of the dominant $H_1$ feature from secondary loops. 
The birth--death interval of the most persistent $H_1$ feature is used to define the filtration scale range for the denoising experiments (Figs.~\ref{fig:sphere_vs_ellipsoid_mean_avg_denoising} and ~\ref{fig:sphere_vs_ellipsoid_denoising_geomedian}).
}
  \label{fig:acc_PD}
\end{figure}

\begin{figure}[t]
  \centering

  \begin{subfigure}{0.9\linewidth}
    \centering
    \includegraphics[width=.9\linewidth]{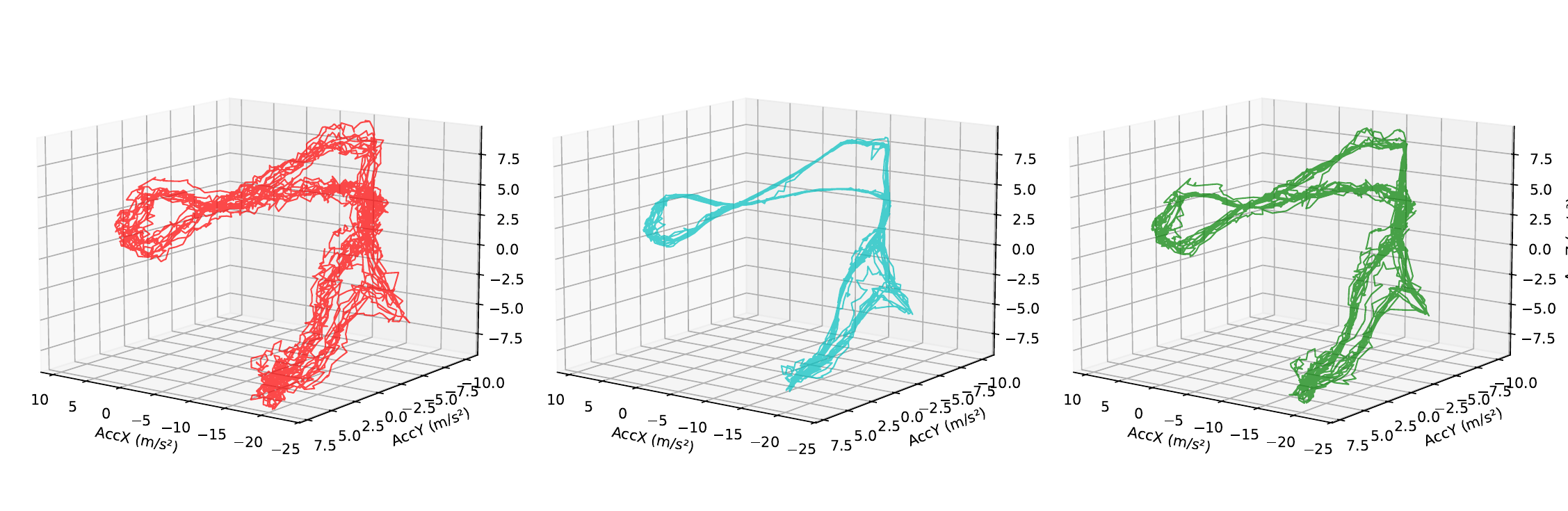}
    %\subcaption{$s=2.494,\ e=2.052$}
  \end{subfigure}

  \vspace{0.5em}

  \begin{subfigure}{0.9\linewidth}
    \centering
    \includegraphics[width=.9\linewidth]{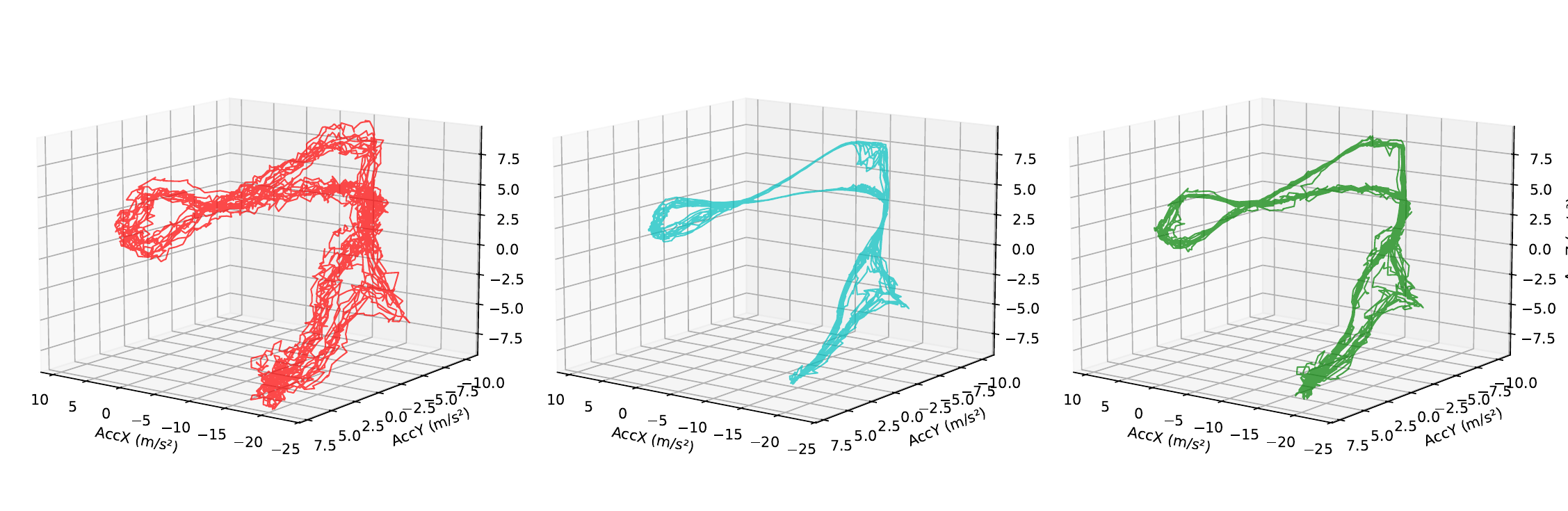}
    %\subcaption{$s=3.626,\ e=3.334$}
  \end{subfigure}

  \vspace{0.5em}

  \begin{subfigure}{0.9\linewidth}
    \centering
    \includegraphics[width=.9\linewidth]{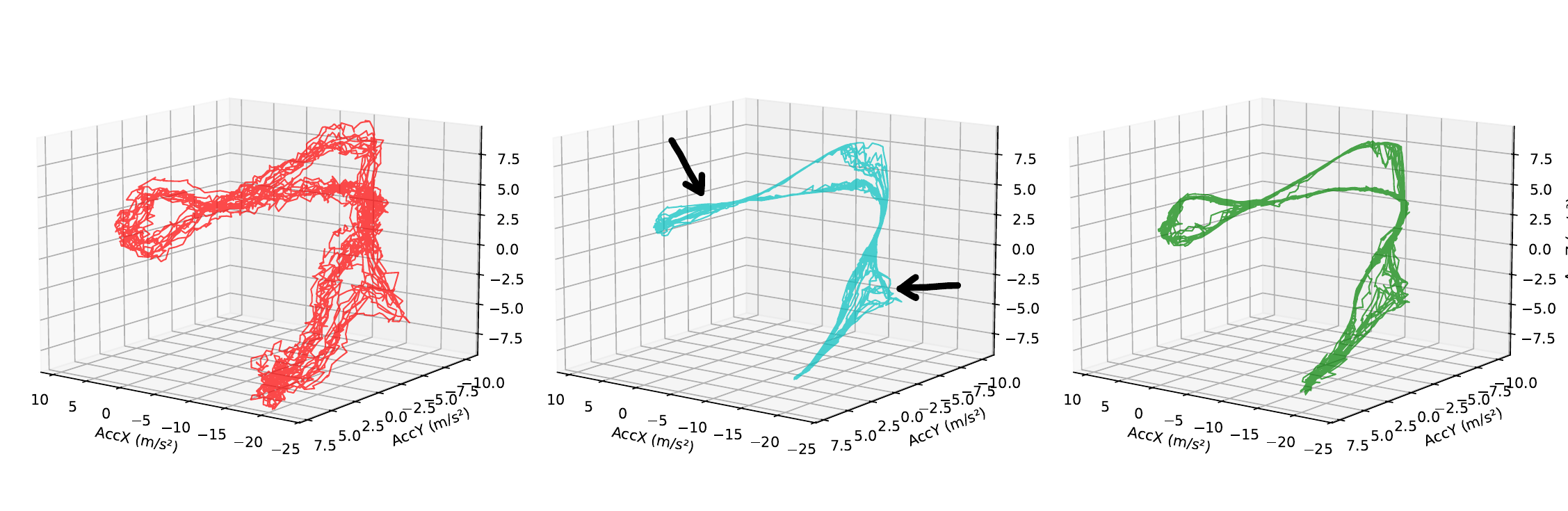}
    %\subcaption{$s=4.759,\ e=4.616$}
  \end{subfigure}

  \vspace{0.5em}

  \begin{subfigure}{0.9\linewidth}
    \centering
    \includegraphics[width=.9\linewidth]{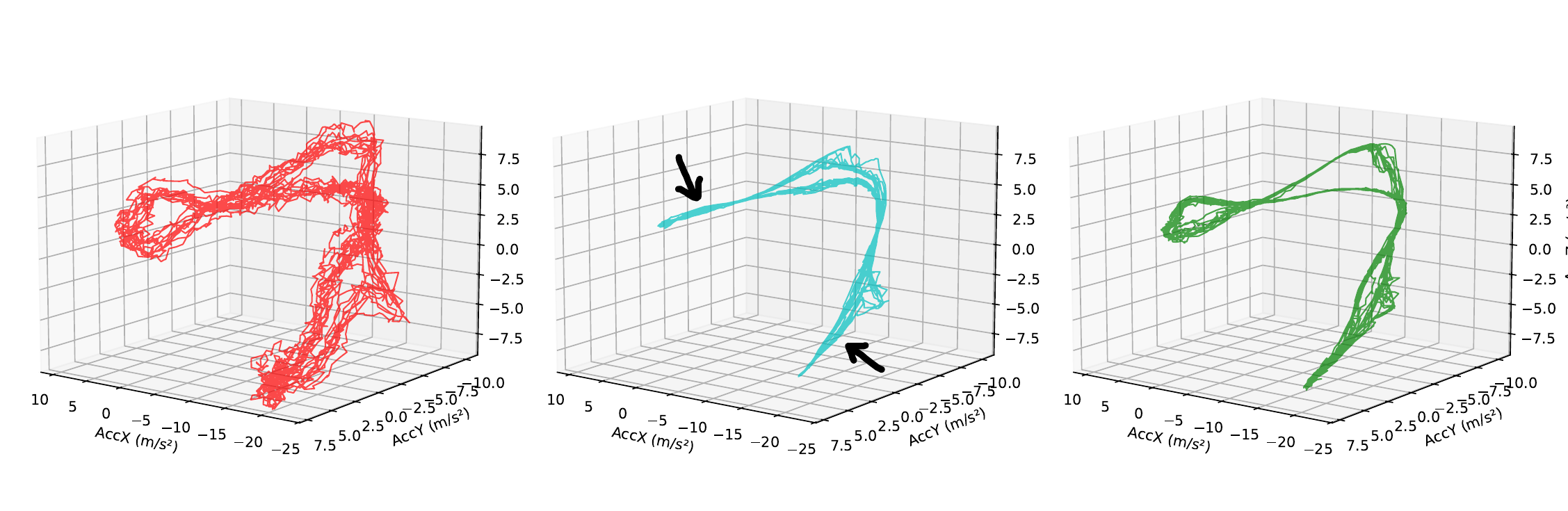}
    %\subcaption{$s=5.891,\ e=5.898$}
  \end{subfigure}

  \caption{
Accelerometer denoising by neighbourhood averaging in state space, comparing isotropic (Vietoris--Rips) and flow-aware (ellipsoidal) neighbourhoods across persistence-normalised scales. 
Each row shows (left) the raw trajectory, (middle) mean-averaging over spherical neighbourhoods, and (right) mean-averaging over ellipsoidal neighbourhoods.
Scales are selected relative to the dominant $H_1$ persistence interval $[B,D]$ of each filtration. 
With $L=D-B$, rows (top to bottom) correspond to $B$, $B+\tfrac{1}{2}L$, $D$, and $D+\tfrac{1}{2}L$.
As the neighbourhood scale increases, progressive tightening and eventual collapse of the loop structure are observed, more pronounced under spherical neighbourhoods. 
Geometric-median results are provided in Appendix~\ref{sec:appendix_geomedian}, Fig.~\ref{fig:sphere_vs_ellipsoid_denoising_geomedian}.
}
  \label{fig:sphere_vs_ellipsoid_mean_avg_denoising}
\end{figure}

We now evaluate the method on real data: publicly available tri-axial accelerometer recordings from cycling \citep{vietenKinematicsCyclicHuman2020}. The data were sampled at 500~Hz; we analysed a 3{,}500-sample segment from the first participant.

To set comparable neighbourhood scales for Vietoris–Rips (VR) and the ellipsoidal filtration, we first computed persistent homology on a shorter 350-sample segment (about one full cycle). Let \(B\) and \(D\) be the birth and death of the dominant \(H_1\) class, and \(L=D-B\) its lifetime. For each filtration we examined four scales: \(B\), \(B+\tfrac{1}{2}L\), \(D\), and \(D+\tfrac{1}{2}L\). When estimating ellipsoidal covariances, we pooled neighbours across adjacent cycles to stabilise local variance estimates, as described in the Methods.

Fig.~\ref{fig:acc_PD} shows persistence diagrams for the same 3D segment under the Vietoris-Rips and the proposed Ellipsoidal filtrations (red: \(H_0\), blue: \(H_1\)). In the VR diagram, two \(H_1\) classes are clearly more persistent, whereas in the ellipsoidal diagram a single \(H_1\) class dominates. The blue points farthest from the diagonal represent the most persistent loop in each panel. In the denoising and recurrence time estimation (Section~\ref{sec:recurrence-time}), we consider the scale of the most persistent \(H_1\) class.

Fig.~\ref{fig:sphere_vs_ellipsoid_mean_avg_denoising} compares spherical and ellipsoidal neighbourhood averaging across four scales. At small scales, both methods preserve the data's loop structure. As the scale increases, spherical neighbourhoods span opposite sides of narrow passages and average across bottlenecks, compressing the geometry and attenuating amplitude. In contrast, ellipsoidal averaging follows the local flow and better preserves loops across scales.

Fig.~\ref{fig:sphere_vs_ellipsoid_denoising_geomedian} applies denoising by taking the geometric median of the samples within each sphere or ellipsoid. The geometric median yields smoother reconstructions and exhibits greater robustness to noise than the mean.

\subsection{Recurrence-time estimation}
\label{sec:recurrence-time}

We assess whether the same flow-aware neighbourhoods that improved denoising (Sections~\ref{sec:DenoisingRecurrentSignal}--\ref{sec:accDenoising}) also improve recurrence-time estimation. The concept of recurrence and first-return times was introduced in Section~\ref{sec:background-recurrence}. For consistency, we use the chirp signal defined in Section~\ref{sec:DenoisingRecurrentSignal}, which combines a linearly increasing frequency, anisotropic deformation, and a narrow bottleneck as described earlier.

In the experimental setting, we consider a discrete sequence of noisy samples \(\{x_i\}_{i=1}^n \subset \mathbb{R}^d\). At each index \(i\), we define a neighbourhood \(\mathcal{N}_i \subset \mathbb{R}^d\) either as a closed Euclidean ball 
\begin{equation}
\mathcal{N}_i = \{\,x \in \mathbb{R}^d : \|x-x_i\|_2 \leq \varepsilon \,\}
\end{equation}
(spherical case), or as a locally aligned ellipsoid 
\begin{equation}
    \mathcal{N}_i = \{\,x \in \mathbb{R}^d : (x-x_i)^\top A_i (x-x_i) \leq 1 \,\},
\end{equation}
where \(A_i \in \mathbb{R}^{d \times d}\) is a positive-definite matrix estimated from local PCA (ellipsoidal case).

The first recurrence time at index \(i\) is then defined as
\begin{equation}
\label{eq:first_recurrence}
\begin{aligned}
T_1(i) = \min \{\, j>i :\; & j-i \ge \tau_{\min},\ x_j \in \mathcal{N}_i,\\
                          & x_k \notin \mathcal{N}_i \ \text{for } i<k<j \}.
\end{aligned}
\end{equation}

where \(\tau_{\min}=15\) samples prevents spurious early returns introduced by noise or oversampling. For reference, ground-truth first returns are computed from the unwrapped phase \(\theta(t)\) as the smallest \(j>i\) such that \(\theta[j] \ge \theta[i] + 2\pi\). As in Section~\ref{sec:accDenoising}, we evaluate four scales per filtration taken from the dominant \(H_1\) class: \(B\), \(B+\tfrac{1}{2}(D-B)\), \(D\), and \(D+\tfrac{1}{2}(D-B)\).

\begin{figure}[t]
  \centering
  % --- TOP ROW ---
  \begin{subfigure}{0.40\linewidth}
    \centering
    \includegraphics[width=\linewidth]{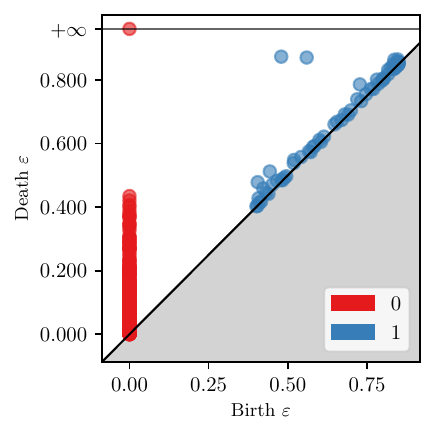}
  \end{subfigure}
  \begin{subfigure}{0.40\linewidth}
    \centering
    \includegraphics[width=\linewidth]{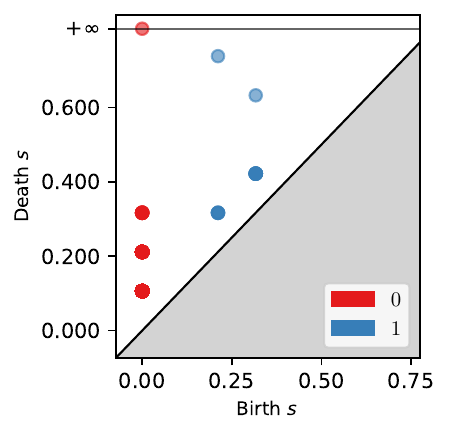}
  \end{subfigure}

  \vspace{0.7em}

  % --- BOTTOM ROW ---
  \begin{subfigure}{0.40\linewidth}
    \centering
    \includegraphics[width=\linewidth]{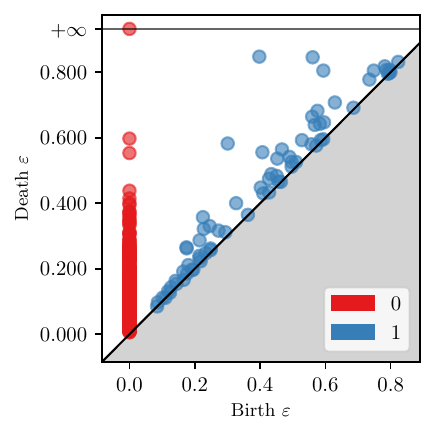}
  \end{subfigure}
  \begin{subfigure}{0.40\linewidth}
    \centering
    \includegraphics[width=\linewidth]{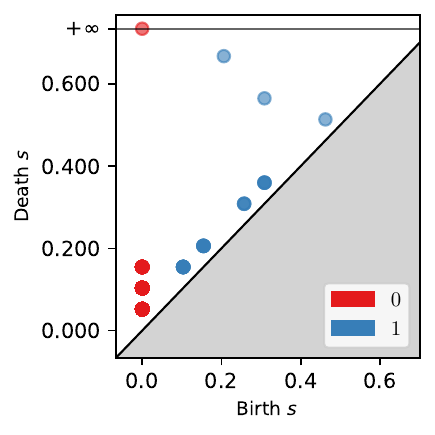}
  \end{subfigure}

  \caption{
Persistent diagrams for clean (top row) and noisy (bottom row) signals under Vietoris--Rips filtration (left column) and flow-aware ellipsoidal filtration (right column). 
Noise introduces additional low-persistence $H_1$ features, visible as points closer to the diagonal.
In each diagram, the $H_1$ point farthest from the diagonal corresponds to the most persistent loop. 
Its birth--death interval defines the dominant $H_1$ persistence range used to select the filtration scales for recurrence-time estimation in Fig.~\ref{fig:recurrence_clean_2x4}.
}
  \label{fig:pd_clean_noisy}
\end{figure}

Fig.~\ref{fig:pd_clean_noisy} presents the persistence diagrams of the clean and noisy signals for both Vietoris–Rips and ellipsoidal filtrations, highlighting the dominant \(H_1\) features under each setting. 

\begin{figure*}[t]
  \centering
  \begingroup
  \renewcommand\thesubfigure{\alph{subfigure}}

  % ================= (A) CLEAN =================
  \begin{subfigure}[t]{\textwidth}
    \centering

    % Top row: Ellipsoids
    \begin{subfigure}{0.245\linewidth}
      \centering
      \includegraphics[width=\linewidth]{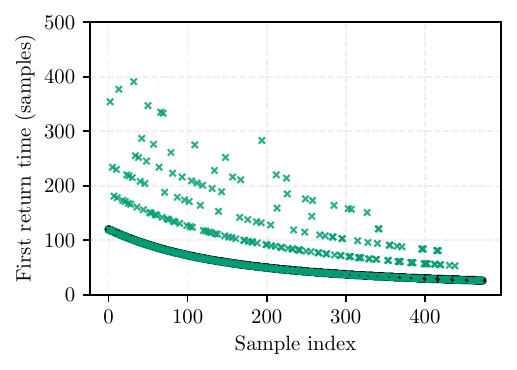}
      %\subcaption*{$\varepsilon=0.211$}
    \end{subfigure}\hfill
    \begin{subfigure}{0.245\linewidth}
      \centering
      \includegraphics[width=\linewidth]{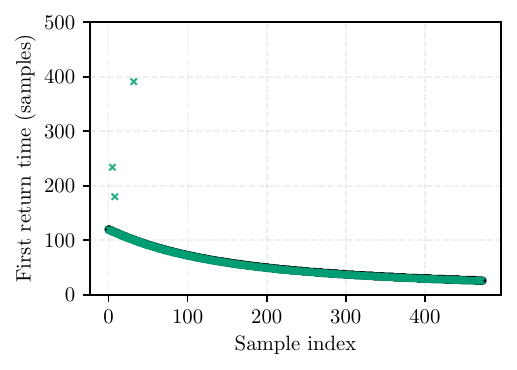}
      %\subcaption*{$\varepsilon=0.580$}
    \end{subfigure}\hfill
    \begin{subfigure}{0.245\linewidth}
      \centering
      \includegraphics[width=\linewidth]{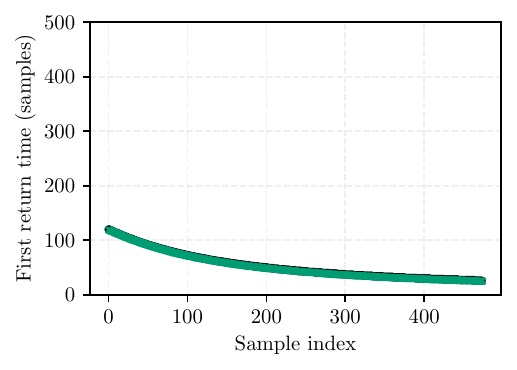}
      %\subcaption*{$\varepsilon=0.948$}
    \end{subfigure}\hfill
    \begin{subfigure}{0.245\linewidth}
      \centering
      \includegraphics[width=\linewidth]{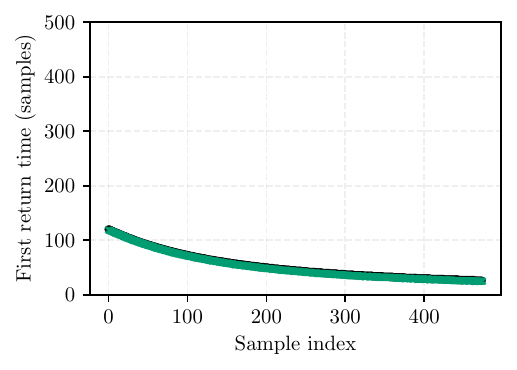}
      %\subcaption*{$\varepsilon=1.316$}
    \end{subfigure}

    \medskip

    % Bottom row: VR
    \begin{subfigure}{0.245\linewidth}
      \centering
      \includegraphics[width=\linewidth]{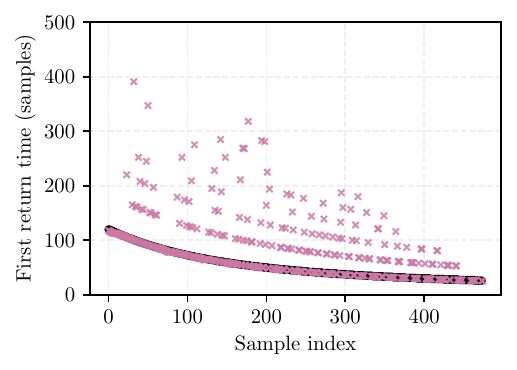}
      %\subcaption*{$r=0.240$}
    \end{subfigure}\hfill
    \begin{subfigure}{0.245\linewidth}
      \centering
      \includegraphics[width=\linewidth]{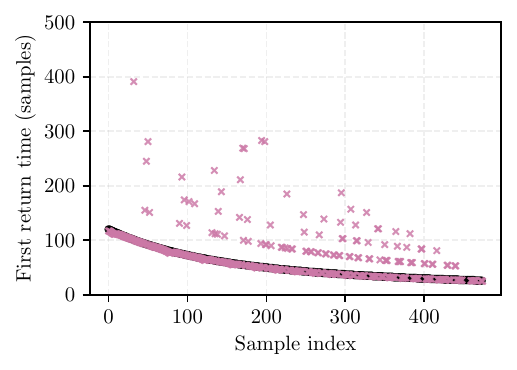}
      %\subcaption*{$r=0.338$}
    \end{subfigure}\hfill
    \begin{subfigure}{0.245\linewidth}
      \centering
      \includegraphics[width=\linewidth]{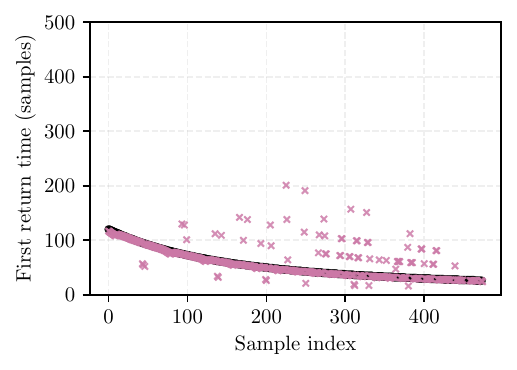}
      %\subcaption*{$r=0.437$}
    \end{subfigure}\hfill
    \begin{subfigure}{0.245\linewidth}
      \centering
      \includegraphics[width=\linewidth]{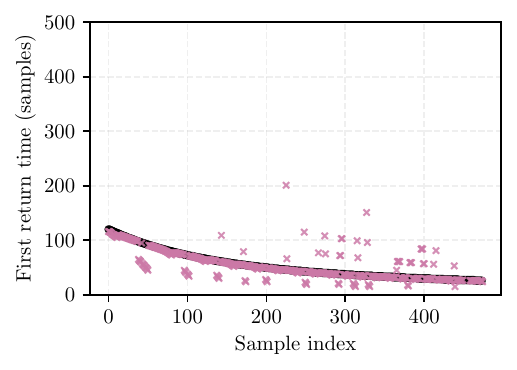}
      %\subcaption*{$r=0.535$}
    \end{subfigure}

    \subcaption{}
    \label{fig:recurrence_clean_2x4}
  \end{subfigure}

  \vspace{0.8em}

  % ================= (B) NOISY =================
  \begin{subfigure}[t]{\textwidth}
    \centering

    % Top row: Ellipsoids
    \begin{subfigure}{0.245\linewidth}
      \centering
      \includegraphics[width=\linewidth]{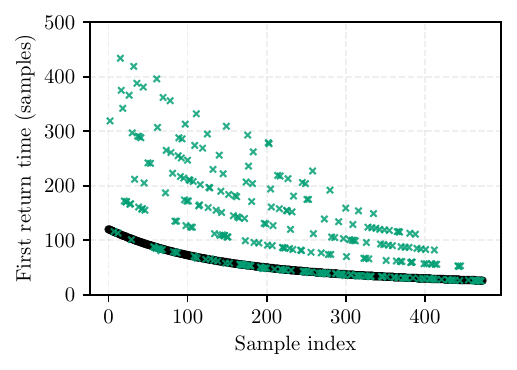}
      %\subcaption*{$\varepsilon=0.211$}
    \end{subfigure}\hfill
    \begin{subfigure}{0.245\linewidth}
      \centering
      \includegraphics[width=\linewidth]{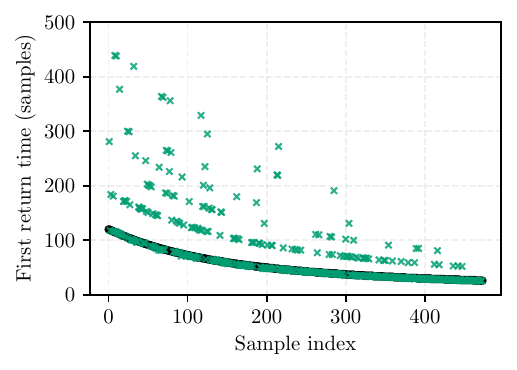}
      %\subcaption*{$\varepsilon=0.527$}
    \end{subfigure}\hfill
    \begin{subfigure}{0.245\linewidth}
      \centering
      \includegraphics[width=\linewidth]{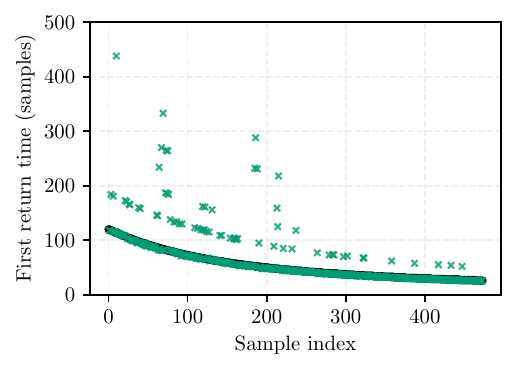}
      %\subcaption*{$\varepsilon=0.843$}
    \end{subfigure}\hfill
    \begin{subfigure}{0.245\linewidth}
      \centering
      \includegraphics[width=\linewidth]{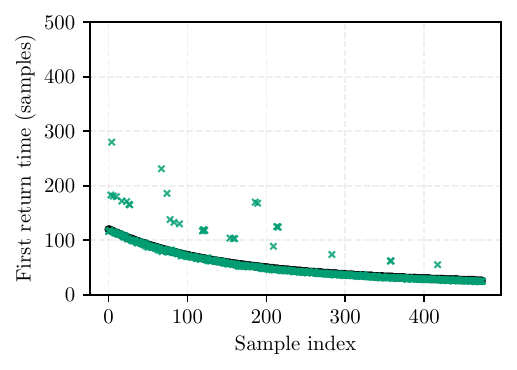}
      %\subcaption*{$\varepsilon=1.158$}
    \end{subfigure}

    \medskip

    % Bottom row: VR
    \begin{subfigure}{0.245\linewidth}
      \centering
      \includegraphics[width=\linewidth]{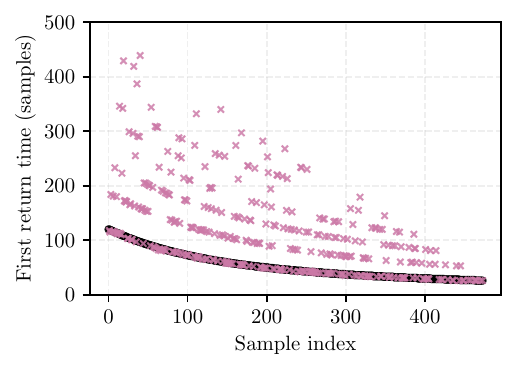}
      %\subcaption*{$r=0.199$}
    \end{subfigure}\hfill
    \begin{subfigure}{0.245\linewidth}
      \centering
      \includegraphics[width=\linewidth]{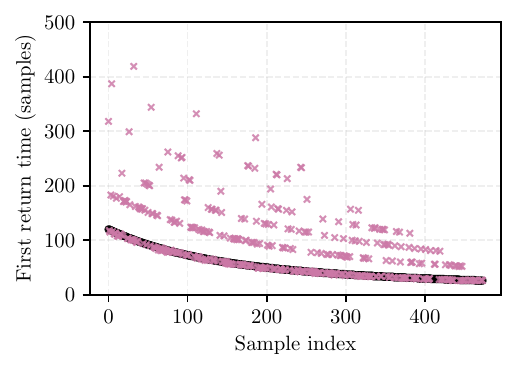}
      %\subcaption*{$r=0.311$}
    \end{subfigure}\hfill
    \begin{subfigure}{0.245\linewidth}
      \centering
      \includegraphics[width=\linewidth]{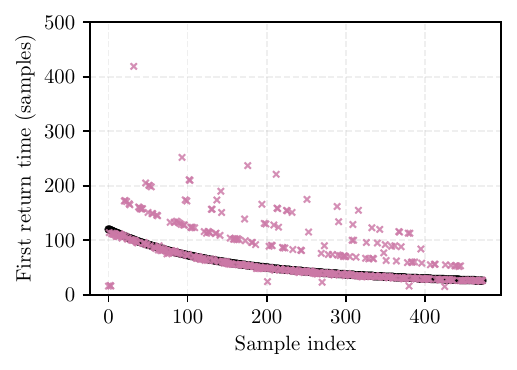}
      %\subcaption*{$r=0.424$}
    \end{subfigure}\hfill
    \begin{subfigure}{0.245\linewidth}
      \centering
      \includegraphics[width=\linewidth]{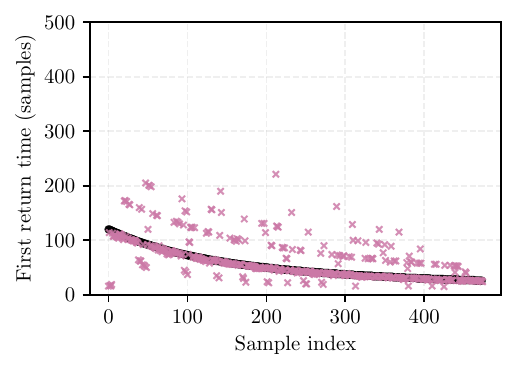}
      %\subcaption*{$r=0.536$}
    \end{subfigure}

    \subcaption{}
    \label{fig:recurrence_noisy_2x4}
  \end{subfigure}

  \endgroup

  \caption{
Recurrence-time estimation for clean (a) and noisy (b) signals using filtration-based recurrence thresholds defined via flow-aware ellipsoidal (anisotropic, top rows) and isotropic Vietoris--Rips (bottom rows) constructions. 
Coloured markers indicate the estimated first return times under the corresponding filtration. 
Alignment of the markers with the black curve indicates accurate estimation. 
Markers substantially below the curve correspond to spurious early returns, typically caused by the narrow pinch region of the trajectory or by noise, whereas markers above the curve indicate returns detected in subsequent cycles rather than the first true recurrence.
Across columns (left to right), the panels correspond to increasing filtration scales from the dominant $H_1$ persistence interval shown in Fig.~\ref{fig:pd_clean_noisy}, namely 
$B$, $B+\tfrac{1}{2}(D-B)$, $D$, and $D+\tfrac{1}{2}(D-B)$, enabling consistent comparison between the two filtrations.
}

  \label{fig:recurrence_clean_noisy_side_by_side}
\end{figure*}

Figs.~\ref{fig:recurrence_clean_2x4} and \ref{fig:recurrence_noisy_2x4} compare first-return-time estimates across persistence-normalised scales. On the clean signal, ellipsoidal neighbourhoods match the ground truth most closely, and at the $H_1$ death scale recover the first return for nearly all indices. In contrast, fixed-radius balls struggle as the period shortens: first returns are frequently missed at all scales, and at larger scales balls span opposite sides of the bottleneck, producing spurious early returns. Under noise, ellipsoidal neighbourhoods capture more first returns. Importantly, no spurious early returns are caused by the bottleneck which makes it possible to estimate returns for all points.

Because the ellipsoidal neighbourhoods are point-specific, the induced recurrence relation is generally asymmetric (i.e.\ $x_j \in \mathcal{N}_i$ does not imply $x_i \in \mathcal{N}_j$), which naturally leads to a directed recurrence plot. In this work, we focus on first-return times and leave the construction of directed recurrence plots, and the corresponding Recurrence Quantification Analysis (RQA) measures, to future work.


\section{Discussion}
\label{sec:discussion}
Across all experiments, adaptive ellipsoidal neighbourhoods preserved the trajectory geometry more consistently than the Vietoris--Rips and Fermat filtrations.  In the Hamiltonian system, they recovered the structure at smaller scales without prematurely bridging the narrow bottleneck, whereas the alternative filtrations connected dense regions too early (Fig.~\ref{fig:filtration_comparison}).  In the synthetic chirp, they retained both amplitude and waveform through the bottleneck and yielded the most accurate reconstruction, particularly for the small-amplitude component $y(t)$ (Figs.~\ref{fig:filtered_signals} and \ref{fig:snr_rmse}).  For the three-dimensional accelerometer data, loop structure persisted across scales under ellipsoidal neighbourhoods, while spherical averaging collapsed the geometry at larger scales (Figs.~\ref{fig:acc_PD} and \ref{fig:sphere_vs_ellipsoid_mean_avg_denoising}).

This behaviour is consistent with an anisotropic neighbourhood model that adapts to local flow variability.  By allowing the neighbourhood shape and orientation to vary with local geometry, the ellipsoidal construction can follow narrow passages without joining distinct regions too early.  However, the current method does not explicitly use tangent vectors when selecting neighbours.  Incorporating tangent-direction information into neighbour selection and/or complex construction may further improve noise robustness.

\begin{figure}[t]
    \centering
    \begin{subfigure}[b]{0.45\linewidth}
        \includegraphics[width=\linewidth]{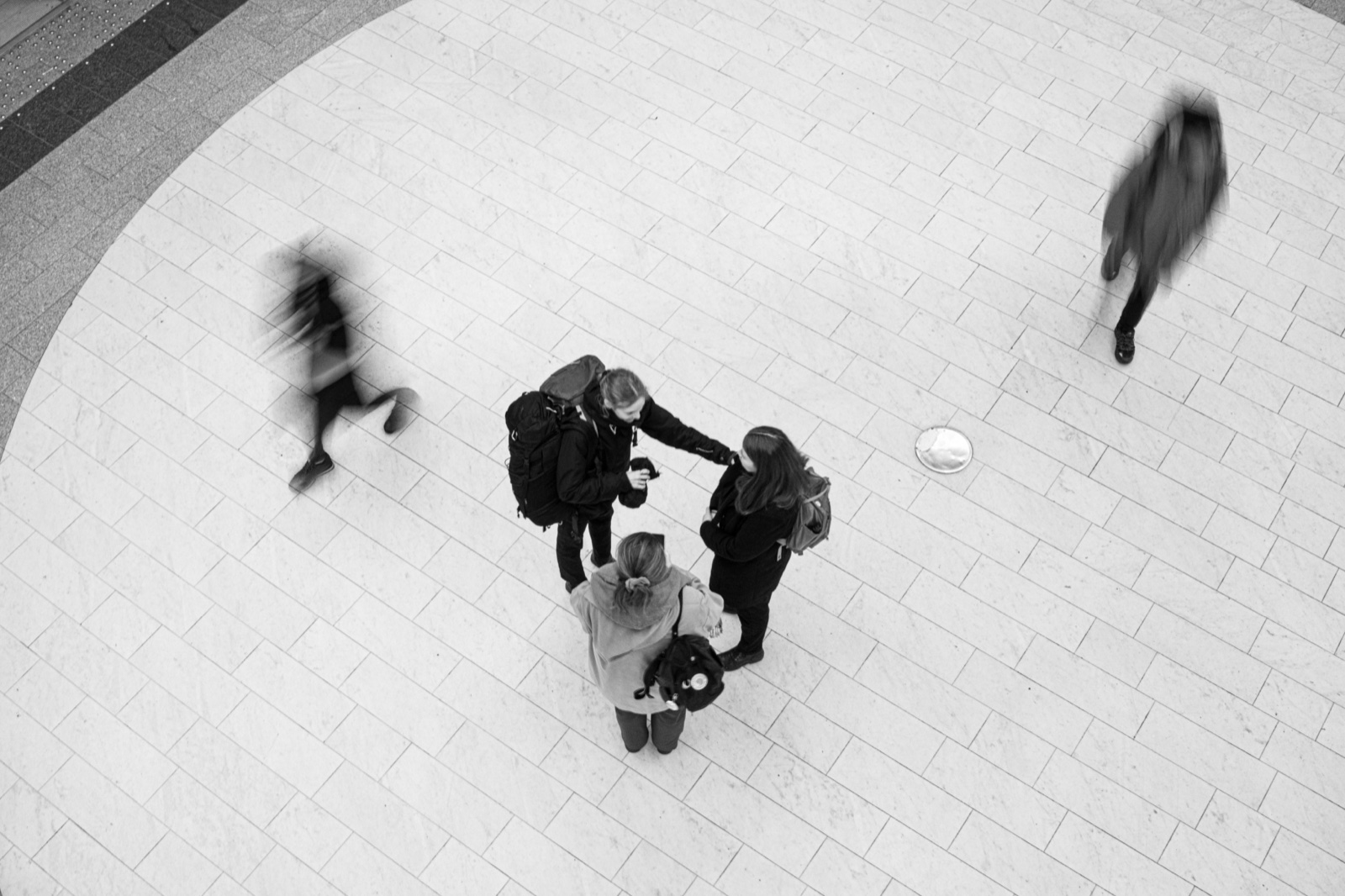}
        %\caption{Motion blur with temporal structure.}
        \label{fig:motion_blur_temporal}
    \end{subfigure}
    \hfill
    \begin{subfigure}[b]{0.45\linewidth}
        \includegraphics[width=\linewidth]{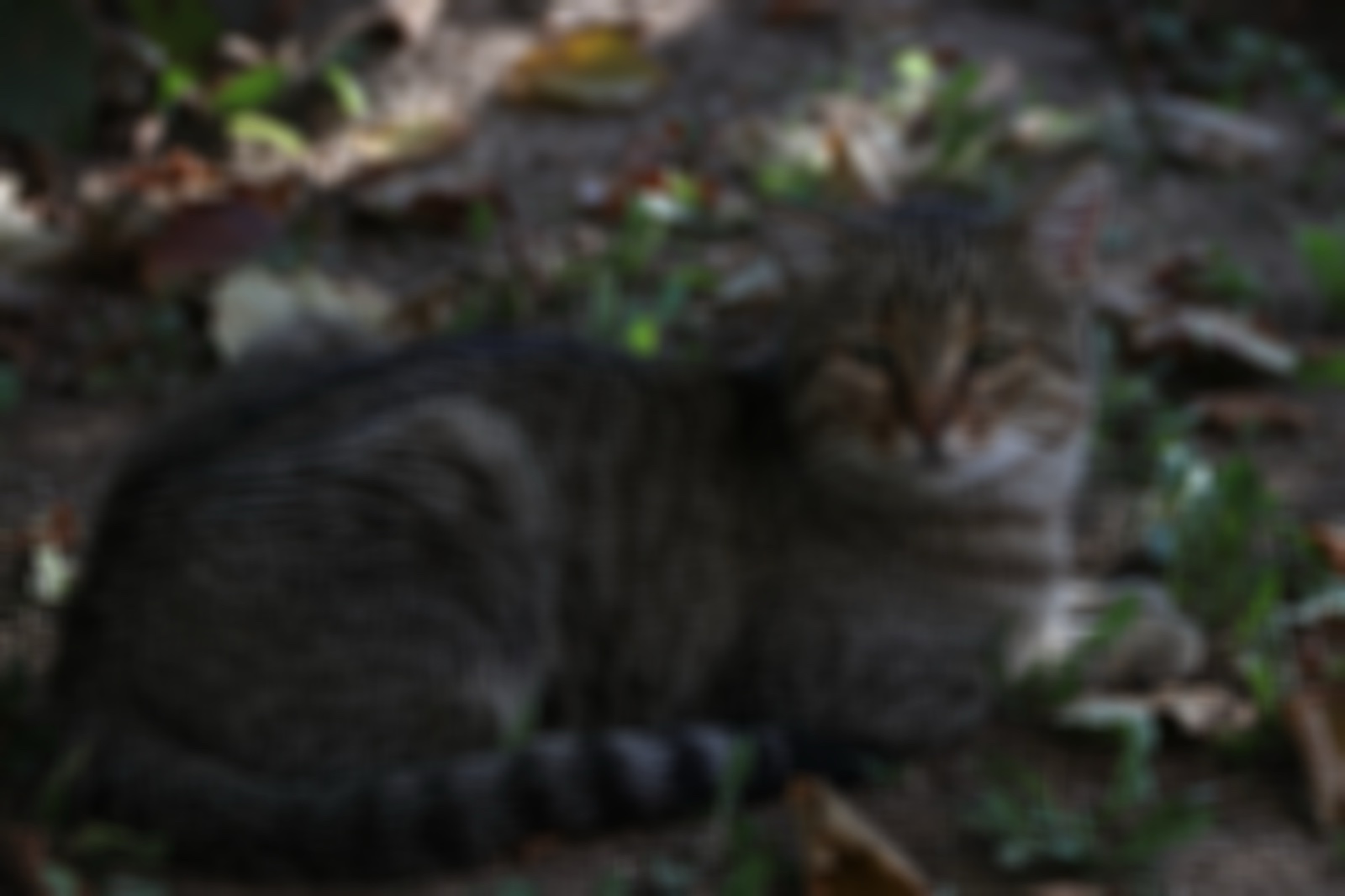}
        %\caption{Blur without temporal awareness.}
        \label{fig:blur_without_temporal}
    \end{subfigure}
    \caption{Analogy between ellipsoidal filtration and motion blur photography.
    Left: motion blur with directional structure.
    Right: blur without directional structure.
    Image credits: Matti Sulanto (left) and İnsunur Eryılmaz (right), used with permission.}
    \label{fig:blur_analogy}
\end{figure}

A complementary interpretation is to view the scale parameter as controlling the size of an uncertainty region around each sample.  As scale increases, these regions expand and the filtration examines how separability changes: when two regions intersect, the corresponding samples are assumed to be no longer distinguishable at that resolution and are connected via edges and higher-dimensional simplices.  The Vietoris--Rips filtration uses isotropic (spherical) uncertainty regions that grow uniformly with scale, whereas the ellipsoidal filtration uses anisotropic (ellipsoidal) regions whose shape is determined by local covariance and tangent-direction estimates from spatio--temporal neighbourhoods.  In dynamical terms, a tangent (velocity) vector encodes change with respect to time, so shaping uncertainty regions using tangent information makes the neighbourhood model temporally informed, even though the filtration parameter remains a geometric scale.  This contrast is analogous to spatial blur versus motion blur: isotropic growth resembles reducing spatial resolution, while anisotropic, direction-aware growth reflects temporal structure (Fig.~\ref{fig:blur_analogy}).

For denoising (Section~\ref{sec:DenoisingRecurrentSignal}), we varied neighbourhoods by both cardinality (\(k\)-nearest neighbours) and geometry (spherical versus ellipsoidal).  Performance improved with increasing SNR for all topological filters, but ellipsoidal neighbourhoods better preserved amplitude and phase near the bottleneck and achieved lower RMSE, especially for $y(t)$ (Figs.~\ref{fig:filtered_signals} and \ref{fig:snr_rmse}).  In contrast, for horizontally elongated data the Vietoris--Rips filtration could promote an incorrect hole to be the most persistent \(H_1\), and increasing noise could inflate that persistence, making the resulting scale unreliable.  In our experiments, the ellipsoidal construction was less affected because the added noise followed the data’s anisotropy, and the method adapts neighbourhood shape using local variance information.  However, these topological filters remain effective only when the data retain a distinguishable structure; strong noise or outliers can obscure the geometry and lead to unreliable features.

For recurrence-time estimation (Section~\ref{sec:recurrence-time}), using the most persistent \(H_1\) class to select scales links the recurrence threshold to the geometry of the trajectory rather than to an arbitrary radius.  Recurrence times were then computed from first-return intervals (Section~\ref{sec:background-recurrence}).  With these scales, flow-aligned neighbourhoods detected first returns under strong frequency drift while avoiding early returns across the bottleneck (Figs.~\ref{fig:recurrence_clean_2x4} and \ref{fig:recurrence_noisy_2x4}).  Temporal neighbours helped the neighbourhoods track changes in the non-autonomous flow.  Additional noise made the \(H_1\)-based scale less reliable for Vietoris--Rips because \(H_1\) persistence increased, whereas the ellipsoidal filtration remained more stable under the same conditions.

A key limitation is sensitivity to outliers: both Vietoris--Rips and ellipsoidal filtrations can produce spurious loops at low \(\mathrm{SNR}\).  The real three-dimensional accelerometer data (Fig.~\ref{fig:sphere_vs_ellipsoid_mean_avg_denoising}) illustrate a noise regime in which the dominant \(H_1\) can guide denoising or recurrence analysis.  In our high-SNR experiments with finite sample lengths, extreme excursions were uncommon and the dominant \(H_1\) provided a stable guide.  At lower SNR, persistence can shrink towards the diagonal or be replaced by spurious loops, so selecting a denoising or recurrence threshold from the most persistent \(H_1\) should be treated with caution and complemented by additional checks.

Another limitation concerns the estimation of tangent vectors for constructing ellipsoids in higher-dimensions. As the ambient dimension increases, the number of samples needed for stable estimation grows rapidly, rendering the present implementation impractical in high dimensions. Moreover, the estimated tangent directions depend on the choice of neighbourhood size and temporal window, so different parameter choices may yield different ellipsoid orientations and reduce reproducibility across datasets. Diffusion‐geometry methods offer alternative strategies for local tangent estimation without assuming a global manifold and without reliance on explicit neighbourhood parameters, and integrating such techniques may improve robustness while preserving the geometric intent of the proposed construction \citep{jones2026manifolddiffusiongeometrycurvature}.

Several directions remain for future work. Incorporating temporal information into diffusion geometry methods might increase the robustness of tangent vector estimation and reduce the number of samples required. Additionally, incorporating tangent-direction information into neighbour selection and the construction of simplicial complexes could enhance robustness. Robust persistent homology methods could be adapted for denoising \citep{anai_dtm-based_2020,damrichPersistentHomologyHighdimensional} and may help mitigate sensitivity to outliers. Finally, persistent-homology-guided scale selection could be explored using alternative similarity measures from recurrence analysis to derive more principled recurrence thresholds \citep{marwanTrendsRecurrenceAnalysis2023}.

\section{Conclusion}
\label{sec:conclusion_and_future_work}
We propose a flow-aware ellipsoidal filtration that treats the observed point cloud as samples from a smooth flow. Using spatio--temporal neighbourhoods and local covariance (PCA), the method fits locally aligned ellipsoids, yielding an adaptive anisotropic filtration. Across our experiments, this construction preserved trajectory geometry through bottlenecks more reliably than isotropic alternatives and provided stable, geometry-informed scales for denoising and recurrence-time estimation.
\begin{acknowledgements}
O.~B.~Eryilmaz acknowledges funding from the Ministry of National Education of the Republic of T\"urkiye.
\end{acknowledgements}
\section*{Author Declarations}

\subsection*{Conflict of Interest}
The authors have no conflicts to disclose.

\appendix
\section{Additional accelerometer denoising results}
\label{sec:appendix_geomedian}

In this appendix we present the accelerometer denoising results obtained using geometric median averaging, complementing the mean-averaging results shown in Fig.~\ref{fig:sphere_vs_ellipsoid_mean_avg_denoising}.

\begin{figure}[htbp]
  \centering

  \begin{subfigure}{0.9\linewidth}
    \centering
    \includegraphics[width=\linewidth]{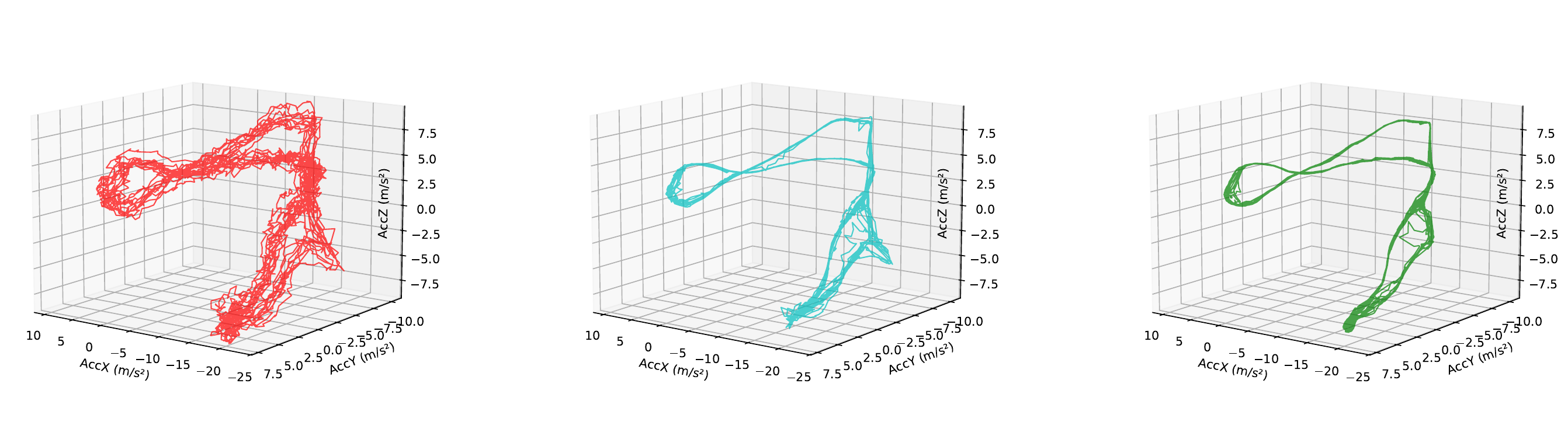}
    %\subcaption{$s=2.494,\ e=1.967$}
  \end{subfigure}

  \vspace{0.6em}

  \begin{subfigure}{0.9\linewidth}
    \centering
    \includegraphics[width=\linewidth]{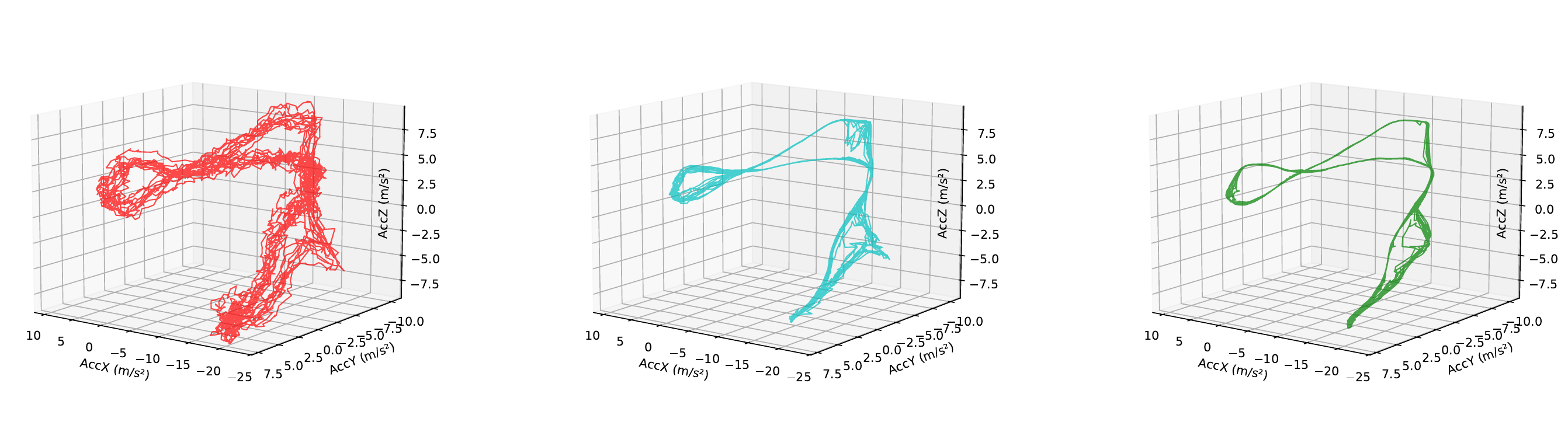}
    %\subcaption{$s=3.626,\ e=3.419$}
  \end{subfigure}

  \vspace{0.6em}

  \begin{subfigure}{0.9\linewidth}
    \centering
    \includegraphics[width=\linewidth]{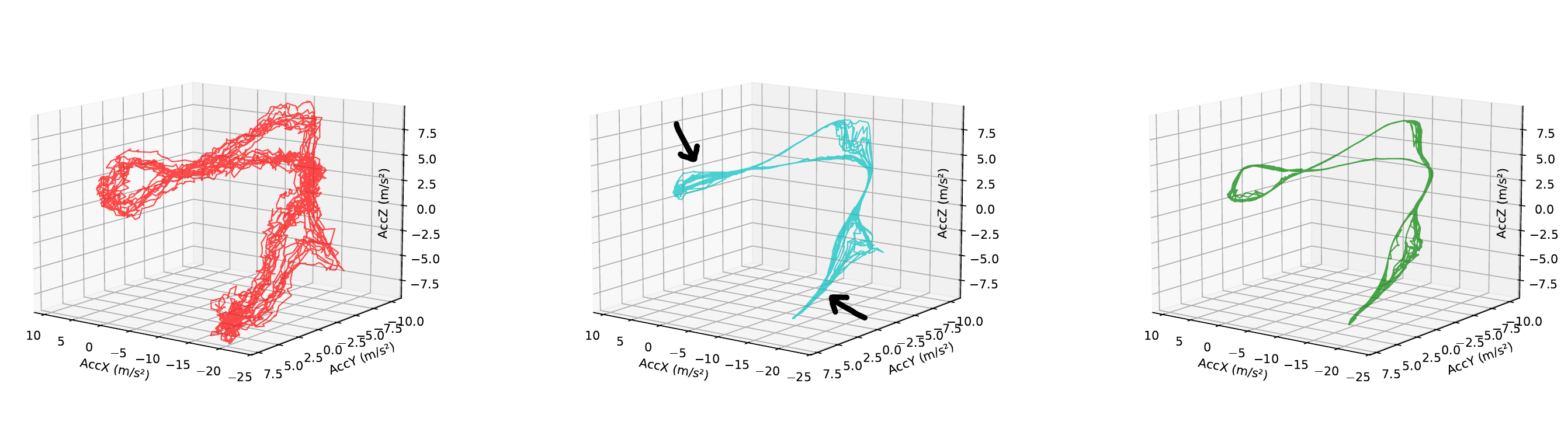}
    %\subcaption{$s=4.759,\ e=4.872$}
  \end{subfigure}

  \vspace{0.6em}

  \begin{subfigure}{0.9\linewidth}
    \centering
    \includegraphics[width=\linewidth]{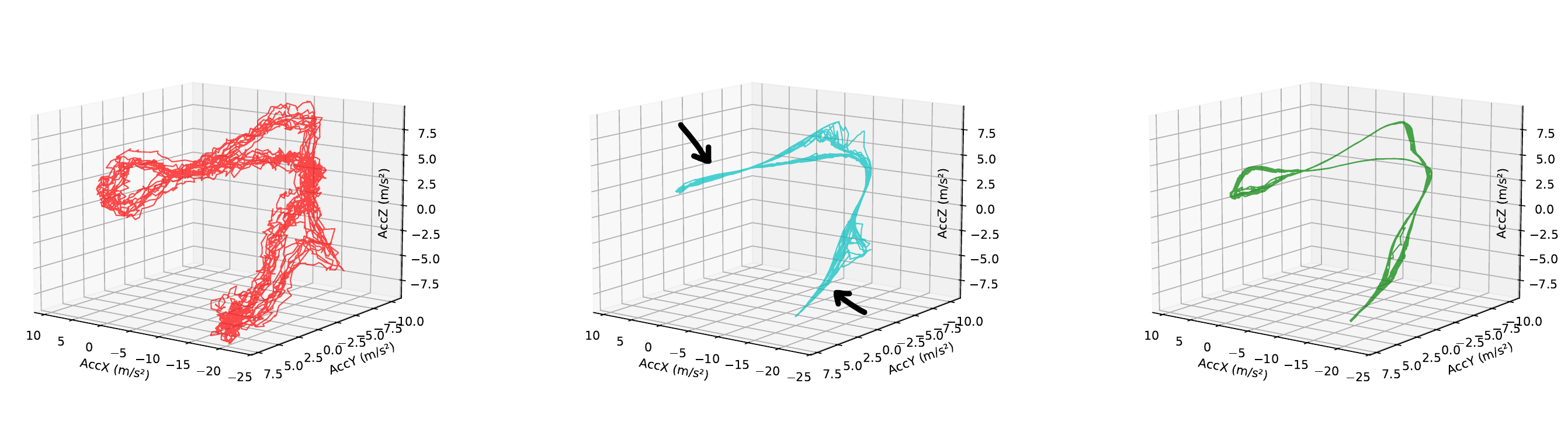}
    %\subcaption{$s=5.891,\ e=6.325$}
  \end{subfigure}

  \caption{
Accelerometer denoising via geometric-median averaging in state space, comparing isotropic (Vietoris--Rips) and flow-aware ellipsoidal neighbourhoods across persistence-normalised scales. 
Each row shows (left) the noisy embedded trajectory, (middle) spherical-neighbourhood denoising, and (right) ellipsoidal-neighbourhood denoising.
Neighbourhood scales are selected relative to the dominant $H_1$ persistence interval $[B,D]$ of each filtration. 
With $L=D-B$, the rows (top to bottom) correspond to $B$, $B+\tfrac{1}{2}L$, $D$, and $D+\tfrac{1}{2}L$, allowing comparison despite differing numerical parameters.
Increasing scale leads to progressive tightening and eventual collapse of the loop structure, which appears earlier under spherical neighbourhoods.
}
  \label{fig:sphere_vs_ellipsoid_denoising_geomedian}
\end{figure}

%\bibliographystyle{plainnat}   
%\bibliography{\mybibfile}      

\begin{thebibliography}{35}
\providecommand{\natexlab}[1]{#1}
\providecommand{\url}[1]{\texttt{#1}}
\expandafter\ifx\csname urlstyle\endcsname\relax
  \providecommand{\doi}[1]{doi: #1}\else
  \providecommand{\doi}{doi: \begingroup \urlstyle{rm}\Url}\fi

\bibitem[Anai et~al.(2020)Anai, Chazal, Glisse, Ike, Inakoshi, Tinarrage, and Umeda]{anai_dtm-based_2020}
Hirokazu Anai, Frédéric Chazal, Marc Glisse, Yuichi Ike, Hiroya Inakoshi, Raphaël Tinarrage, and Yuhei Umeda.
\newblock {DTM}-{Based} {Filtrations}.
\newblock In Nils~A. Baas, Gunnar~E. Carlsson, Gereon Quick, Markus Szymik, and Marius Thaule, editors, \emph{Topological {Data} {Analysis}}, pages 33--66, Cham, 2020. Springer International Publishing.
\newblock ISBN 978-3-030-43408-3.
\newblock \doi{10.1007/978-3-030-43408-3_2}.

\bibitem[Breiding et~al.(2018)Breiding, Kali{\v s}nik, Sturmfels, and Weinstein]{breidingLearningAlgebraicVarieties2018}
Paul Breiding, Sara Kali{\v s}nik, Bernd Sturmfels, and Madeleine Weinstein.
\newblock Learning algebraic varieties from samples.
\newblock \emph{Revista Matem\'atica Complutense}, 31\penalty0 (3):\penalty0 545--593, September 2018.
\newblock ISSN 1988-2807.
\newblock \doi{10.1007/s13163-018-0273-6}.

\bibitem[Canova et~al.(2025)Canova, Kali{\v s}nik, Moser, Rieck, and {\v Z}egarac]{canovaPersistentHomologyEllipsoids2025}
Niklas Canova, Sara Kali{\v s}nik, Aaron Moser, Bastian Rieck, and Ana {\v Z}egarac.
\newblock Persistent {{Homology}} via {{Ellipsoids}}, October 2025.

\bibitem[Carlsson(2009)]{carlssonTopologyData2009}
Gunnar Carlsson.
\newblock Topology and data.
\newblock \emph{Bulletin of the American Mathematical Society}, 46\penalty0 (2):\penalty0 255--308, January 2009.
\newblock ISSN 0273-0979.
\newblock \doi{10.1090/S0273-0979-09-01249-X}.

\bibitem[Carlsson and {Vejdemo-Johansson}(2021)]{carlssonTopologicalDataAnalysis2021}
Gunnar Carlsson and Mikael {Vejdemo-Johansson}.
\newblock \emph{Topological {{Data Analysis}} with {{Applications}}}.
\newblock Cambridge University Press, Cambridge, 2021.
\newblock ISBN 978-1-108-83865-8.
\newblock \doi{10.1017/9781108975704}.

\bibitem[Chazal and Michel(2021)]{chazalIntroductionTopologicalData2021}
Fr{\'e}d{\'e}ric Chazal and Bertrand Michel.
\newblock An {{Introduction}} to {{Topological Data Analysis}}: {{Fundamental}} and {{Practical Aspects}} for {{Data Scientists}}.
\newblock \emph{Frontiers in Artificial Intelligence}, 4, September 2021.
\newblock ISSN 2624-8212.
\newblock \doi{10.3389/frai.2021.667963}.

\bibitem[Damrich et~al.()Damrich, Berens, and Kobak]{damrichPersistentHomologyHighdimensional}
Sebastian Damrich, Philipp Berens, and Dmitry Kobak.
\newblock Persistent {{Homology}} for {{High-dimensional Data Based}} on {{Spectral Methods}}.

\bibitem[{Edelsbrunner} et~al.(2002){Edelsbrunner}, {Letscher}, and {Zomorodian}]{edelsbrunnerTopologicalPersistenceSimplification2002}
{Edelsbrunner}, {Letscher}, and {Zomorodian}.
\newblock Topological {{Persistence}} and {{Simplification}}.
\newblock \emph{Discrete \& Computational Geometry}, 28\penalty0 (4):\penalty0 511--533, November 2002.
\newblock ISSN 1432-0444.
\newblock \doi{10.1007/s00454-002-2885-2}.

\bibitem[Edelsbrunner and Harer(2008)]{Edelsbrunner_Persistent_Homology_a_Survey}
Herbert Edelsbrunner and John Harer.
\newblock Persistent homology---a survey.
\newblock In Jacob~E. Goodman, J{\'a}nos Pach, and Richard Pollack, editors, \emph{Contemporary {{Mathematics}}}, volume 453, pages 257--282. American Mathematical Society, Providence, Rhode Island, 2008.
\newblock ISBN 978-0-8218-4239-3 978-0-8218-8132-3.

\bibitem[Eryilmaz et~al.(2025)Eryilmaz, Katar, and Little]{eryilmazEllipsoidalFiltrationTopological2025}
Omer~Bahadir Eryilmaz, Cihan Katar, and Max~A Little.
\newblock Ellipsoidal {{Filtration}} for {{Topological Denoising}} of {{Recurrent Signals}}.
\newblock In \emph{Proceedings of the 2025 {{International Symposium}} on {{Nonlinear Theory}} and {{Its Applications}} ({{NOLTA2025}})}, pages 610--613, Naha, Okinawa, Japan, October 2025. IEICE.

\bibitem[Fefferman et~al.(2016)Fefferman, Mitter, and Narayanan]{feffermanTestingManifoldHypothesis2016}
Charles Fefferman, Sanjoy Mitter, and Hariharan Narayanan.
\newblock Testing the manifold hypothesis.
\newblock \emph{Journal of the American Mathematical Society}, 29\penalty0 (4):\penalty0 983--1049, October 2016.
\newblock ISSN 0894-0347, 1088-6834.
\newblock \doi{10.1090/jams/852}.

\bibitem[Fern{\'a}ndez et~al.(2023)Fern{\'a}ndez, Borghini, Mindlin, and Groisman]{fernandezIntrinsicPersistentHomology2023}
Ximena Fern{\'a}ndez, Eugenio Borghini, Gabriel Mindlin, and Pablo Groisman.
\newblock Intrinsic {{Persistent Homology}} via {{Density-based Metric Learning}}.
\newblock \emph{Journal of Machine Learning Research}, 24\penalty0 (75):\penalty0 1--42, 2023.
\newblock ISSN 1533-7928.

\bibitem[Gakhar and Perea(2024)]{gakharSlidingWindowPersistence2024}
Hitesh Gakhar and Jose~A. Perea.
\newblock Sliding window persistence of quasiperiodic functions.
\newblock \emph{Journal of Applied and Computational Topology}, 8\penalty0 (1):\penalty0 55--92, March 2024.
\newblock ISSN 2367-1734.
\newblock \doi{10.1007/s41468-023-00136-7}.

\bibitem[Gilitschenski and Hanebeck(2012)]{Gilitschenski_Ellipsoid_Intersection}
Igor Gilitschenski and Uwe~D. Hanebeck.
\newblock A robust computational test for overlap of two arbitrary-dimensional ellipsoids in fault-detection of {{Kalman}} filters.
\newblock In \emph{2012 15th {{International Conference}} on {{Information Fusion}}}, pages 396--401, July 2012.

\bibitem[Giunti et~al.(2022)Giunti, Lazovskis, and Rieck]{DONUT}
Barbara Giunti, J{\=a}nis Lazovskis, and Bastian Rieck.
\newblock {{DONUT}}: {{Database}} of {{Original}} \& {{Non-Theoretical Uses}} of {{Topology}}, 2022.

\bibitem[Hussain~Shah et~al.(2025)Hussain~Shah, Rafia~Fatima, {Huerta-Cuellar}, {Garc{\'i}a-L{\'o}pez}, Mata~Ramirez, and {Jaimes-Re{\'a}tegui}]{hussainshahTopologicalDataAnalysis2025}
W.~Hussain~Shah, S.~Rafia~Fatima, G.~{Huerta-Cuellar}, J.~H. {Garc{\'i}a-L{\'o}pez}, C.~G. Mata~Ramirez, and R.~{Jaimes-Re{\'a}tegui}.
\newblock Topological data analysis approach to time series and shape analysis of dynamical system.
\newblock \emph{Chaos: An Interdisciplinary Journal of Nonlinear Science}, 35\penalty0 (6):\penalty0 063129, June 2025.
\newblock ISSN 1054-1500.
\newblock \doi{10.1063/5.0268340}.

\bibitem[Jones(2026)]{jones2026manifolddiffusiongeometrycurvature}
Iolo Jones.
\newblock Manifold diffusion geometry: Curvature, tangent spaces, and dimension, 2026.
\newblock URL \url{https://arxiv.org/abs/2411.04100}.

\bibitem[Kali{\v s}nik and Le{\v s}nik(2024)]{kalisnikFindingHomologyManifolds2024}
Sara Kali{\v s}nik and Davorin Le{\v s}nik.
\newblock Finding the homology of manifolds using ellipsoids.
\newblock \emph{Journal of Applied and Computational Topology}, 8\penalty0 (1):\penalty0 193--238, March 2024.
\newblock ISSN 2367-1734.
\newblock \doi{10.1007/s41468-023-00145-6}.

\bibitem[Kantz and Schreiber(2003)]{kantzNonlinearTimeSeries2003}
Holger Kantz and Thomas Schreiber.
\newblock \emph{Nonlinear {{Time Series Analysis}}}.
\newblock Cambridge University Press, November 2003.
\newblock ISBN 978-1-139-44043-1.

\bibitem[Li et~al.(2025)Li, Lai, and Oberst]{liOptimizingSelforganizedTopology2025}
Conggai Li, Joseph C.~S. Lai, and Sebastian Oberst.
\newblock Optimizing self-organized topology of recurrence-based complex networks.
\newblock \emph{Chaos: An Interdisciplinary Journal of Nonlinear Science}, 35\penalty0 (3):\penalty0 031101, March 2025.
\newblock ISSN 1054-1500.
\newblock \doi{10.1063/5.0249500}.

\bibitem[Little et~al.(2007)Little, McSharry, Roberts, Costello, and Moroz]{littleExploitingNonlinearRecurrence2007}
Max Little, Patrick McSharry, Stephen Roberts, Declan Costello, and Irene Moroz.
\newblock Exploiting {{Nonlinear Recurrence}} and {{Fractal Scaling Properties}} for {{Voice Disorder Detection}}.
\newblock \emph{Nature Precedings}, pages 1--1, July 2007.
\newblock ISSN 1756-0357.
\newblock \doi{10.1038/npre.2007.326.1}.

\bibitem[Maleti{\'c} et~al.(2016)Maleti{\'c}, Zhao, and Rajkovi{\'c}]{maleticPersistentTopologicalFeatures2016}
Slobodan Maleti{\'c}, Yi~Zhao, and Milan Rajkovi{\'c}.
\newblock Persistent topological features of dynamical systems.
\newblock \emph{Chaos: An Interdisciplinary Journal of Nonlinear Science}, 26\penalty0 (5):\penalty0 053105, May 2016.
\newblock ISSN 1054-1500.
\newblock \doi{10.1063/1.4949472}.

\bibitem[Marchetti(2025)]{marchettiIntrinsicDimensionalityFermi2025}
Gionni Marchetti.
\newblock Intrinsic dimensionality of {{Fermi}}--{{Pasta-Ulam-Tsingou}} high-dimensional trajectories through manifold learning: {{A}} linear approach.
\newblock \emph{Chaos: An Interdisciplinary Journal of Nonlinear Science}, 35\penalty0 (10):\penalty0 103118, October 2025.
\newblock ISSN 1054-1500.
\newblock \doi{10.1063/5.0293702}.

\bibitem[Maria et~al.(2014)Maria, Boissonnat, Glisse, and Yvinec]{mariaGudhiLibrarySimplicial2014}
Cl{\'e}ment Maria, Jean-Daniel Boissonnat, Marc Glisse, and Mariette Yvinec.
\newblock The {{Gudhi Library}}: {{Simplicial Complexes}} and {{Persistent Homology}}.
\newblock In Hoon Hong and Chee Yap, editors, \emph{Mathematical {{Software}} -- {{ICMS}} 2014}, volume 8592, pages 167--174. Springer Berlin Heidelberg, Berlin, Heidelberg, 2014.
\newblock ISBN 978-3-662-44198-5 978-3-662-44199-2.
\newblock \doi{10.1007/978-3-662-44199-2_28}.

\bibitem[Marwan and Kraemer(2023)]{marwanTrendsRecurrenceAnalysis2023}
Norbert Marwan and K.~Hauke Kraemer.
\newblock Trends in recurrence analysis of dynamical systems.
\newblock \emph{The European Physical Journal Special Topics}, 232\penalty0 (1):\penalty0 5--27, February 2023.
\newblock ISSN 1951-6401.
\newblock \doi{10.1140/epjs/s11734-022-00739-8}.

\bibitem[Perea and Harer(2015)]{pereaSlidingWindowsPersistence2015}
Jose~A. Perea and John Harer.
\newblock Sliding {{Windows}} and {{Persistence}}: {{An Application}} of {{Topological Methods}} to {{Signal Analysis}}.
\newblock \emph{Foundations of Computational Mathematics}, 15\penalty0 (3):\penalty0 799--838, June 2015.
\newblock ISSN 1615-3383.
\newblock \doi{10.1007/s10208-014-9206-z}.

\bibitem[Robinson(2016)]{robinsonTopologicalLowPassFilter2016}
Michael Robinson.
\newblock A {{Topological Low-Pass Filter}} for {{Quasi-Periodic Signals}}.
\newblock \emph{IEEE Signal Processing Letters}, 23\penalty0 (12):\penalty0 1771--1775, December 2016.
\newblock ISSN 1070-9908, 1558-2361.
\newblock \doi{10.1109/LSP.2016.2619678}.

\bibitem[Schinkel et~al.(2008)Schinkel, Dimigen, and Marwan]{schinkelSelectionRecurrenceThreshold2008}
S.~Schinkel, O.~Dimigen, and N.~Marwan.
\newblock Selection of recurrence threshold for signal detection.
\newblock \emph{The European Physical Journal Special Topics}, 164\penalty0 (1):\penalty0 45--53, October 2008.
\newblock ISSN 1951-6401.
\newblock \doi{10.1140/epjst/e2008-00833-5}.

\bibitem[Scoccola et~al.(2023)Scoccola, Gakhar, Bush, Schonsheck, Rask, Zhou, and Perea]{scoccolaToroidalCoordinatesDecorrelating2023}
Luis Scoccola, Hitesh Gakhar, Johnathan Bush, Nikolas Schonsheck, Tatum Rask, Ling Zhou, and Jose~A. Perea.
\newblock Toroidal {{Coordinates}}: {{Decorrelating Circular Coordinates}} with {{Lattice Reduction}}.
\newblock \emph{LIPIcs, Volume 258, SoCG 2023}, 258:\penalty0 57:1--57:20, 2023.
\newblock ISSN 1868-8969.
\newblock \doi{10.4230/LIPICS.SOCG.2023.57}.

\bibitem[Tan et~al.(2021)Tan, Corr{\^e}a, Stemler, and Small]{tanGradingYourModels2021}
Eugene Tan, D{\'e}bora Corr{\^e}a, Thomas Stemler, and Michael Small.
\newblock Grading your models: {{Assessing}} dynamics learning of models using persistent homology.
\newblock \emph{Chaos: An Interdisciplinary Journal of Nonlinear Science}, 31\penalty0 (12):\penalty0 123109, December 2021.
\newblock ISSN 1054-1500.
\newblock \doi{10.1063/5.0073722}.

\bibitem[Tan et~al.(2023)Tan, Algar, Corr{\^e}a, Small, Stemler, and Walker]{tanSelectingEmbeddingDelays2023}
Eugene Tan, Shannon Algar, D{\'e}bora Corr{\^e}a, Michael Small, Thomas Stemler, and David Walker.
\newblock Selecting embedding delays: {{An}} overview of embedding techniques and a new method using persistent homology.
\newblock \emph{Chaos: An Interdisciplinary Journal of Nonlinear Science}, 33\penalty0 (3):\penalty0 032101, March 2023.
\newblock ISSN 1054-1500.
\newblock \doi{10.1063/5.0137223}.

\bibitem[{Vejdemo-Johansson} et~al.(2015){Vejdemo-Johansson}, Pokorny, Skraba, and Kragic]{vejdemo-johanssonCohomologicalLearningPeriodic2015}
Mikael {Vejdemo-Johansson}, Florian~T. Pokorny, Primoz Skraba, and Danica Kragic.
\newblock Cohomological learning of periodic motion.
\newblock \emph{Applicable Algebra in Engineering, Communication and Computing}, 26\penalty0 (1):\penalty0 5--26, March 2015.
\newblock ISSN 1432-0622.
\newblock \doi{10.1007/s00200-015-0251-x}.

\bibitem[Vieten and Weich(2020)]{vietenKinematicsCyclicHuman2020}
Manfred~M. Vieten and Christian Weich.
\newblock The kinematics of cyclic human movement.
\newblock \emph{PLoS ONE}, 15\penalty0 (3):\penalty0 e0225157, March 2020.
\newblock ISSN 1932-6203.
\newblock \doi{10.1371/journal.pone.0225157}.

\bibitem[Webber and Marwan(2015)]{webberRecurrenceQuantificationAnalysis2015}
Charles~L. Webber and Norbert Marwan, editors.
\newblock \emph{Recurrence {{Quantification Analysis}}: {{Theory}} and {{Best Practices}}}.
\newblock Understanding {{Complex Systems}}. Springer International Publishing, Cham, 2015.
\newblock ISBN 978-3-319-07154-1 978-3-319-07155-8.
\newblock \doi{10.1007/978-3-319-07155-8}.

\bibitem[Zou et~al.(2019)Zou, Donner, Marwan, Donges, and Kurths]{zouComplexNetworkApproaches2019}
Yong Zou, Reik~V. Donner, Norbert Marwan, Jonathan~F. Donges, and J{\"u}rgen Kurths.
\newblock Complex network approaches to nonlinear time series analysis.
\newblock \emph{Physics Reports}, 787:\penalty0 1--97, January 2019.
\newblock ISSN 0370-1573.
\newblock \doi{10.1016/j.physrep.2018.10.005}.

\end{thebibliography}

\end{document}